\documentclass[preprint2]{aastex}

\newcommand{\2}{{~\sc ii}}

\newcommand{\kms}{{\,km\,s$^{-1}$}}
\newcommand{\mic}{{\,$\mu$m}}

\usepackage{natbib}
\bibpunct{(}{)}{;}{a}{}{,} 
\usepackage{appendix}
\usepackage{float}

\usepackage{graphicx}
\usepackage{transparent}
\usepackage{txfonts}
\usepackage{hyperref}

\begin{document}


\title{CASSIS: The Cornell Atlas of Spitzer/Infrared Spectrograph Sources \\
II.\ High-resolution observations}

   \shorttitle{CASSIS: high-resolution pipeline and atlas}
   	\shortauthors{Lebouteiller et al.}


\author{V.\ Lebouteiller\altaffilmark{1}, D.\ J.\ Barry\altaffilmark{2}, C.\ Goes\altaffilmark{2}, G.\ C.\ Sloan\altaffilmark{2}, H.\ W.\ W. Spoon\altaffilmark{2}, D. W. Weedman\altaffilmark{2}, J.\ Bernard-Salas\altaffilmark{3}, and J.\ R.\ Houck\altaffilmark{2}}

\affil{$^1$ Laboratoire AIM, CEA/DSM-CNRS-Universit\'e Paris Diderot DAPNIA/Service d'Astrophysique B\^at. 709, CEA-Saclay F-91191 Gif-sur-Yvette C\'edex, France}
\affil{$^2$ Department of Astronomy and Center for Radiophysics and Space Research, Cornell} \affil{University, Space Sciences Building, Ithaca, NY 14853-6801, USA\\$^3$ Department of Physical Sciences, The Open University, Milton Keynes, MK7 6AA, UK}








\begin{abstract}
The \textit{Infrared Spectrograph} (IRS) on board the \textit{Spitzer} Space Telescope observed about $15\,000$ objects during the cryogenic mission lifetime. Observations provided low-resolution ($R=\lambda/\Delta\lambda\approx60-127$) spectra over $\approx5-38$\mic\ and high-resolution ($R\approx600$) spectra over $10-37$\mic. The Cornell Atlas of Spitzer/IRS Sources (CASSIS) was created to provide publishable quality spectra to the community. Low-resolution spectra have been available in CASSIS since 2011, and we present here the addition of the high-resolution spectra. 
The high-resolution observations represent approximately one third of all staring observations performed with the IRS instrument. While low-resolution observations are adapted to faint objects and/or broad spectral features (e.g., dust continuum, molecular bands), high-resolution observations allow more accurate measurements of narrow features (e.g., ionic emission lines) as well as a better sampling of the spectral profile of various  features. 
Given the narrow aperture  of the two high-resolution modules, cosmic ray hits and spurious features usually plague the spectra. Our pipeline is designed to minimize these effects through various improvements. A super sampled point-spread function was created in order to enable the optimal extraction in addition to the full aperture extraction. The pipeline selects the best extraction method based on the spatial extent of the object. For unresolved sources, the optimal extraction provides a significant improvement in signal-to-noise ratio over a full aperture extraction. We have developed several techniques for optimal extraction, including a differential method that eliminates low-level rogue pixels (even when no dedicated background observation was performed).  
The updated CASSIS repository now includes all the spectra ever taken by the IRS, with the exception of mapping observations. 
\end{abstract}


   \maketitle

\section{Introduction}

The Infrared Spectrograph (IRS; \citealt{Houck04a})\footnote{The IRS was a collaborative venture between Cornell University and Ball Aerospace Corporation funded by NASA through the
Jet Propulsion Laboratory and the Ames Research Center.} is one of three instruments on board the \textit{Spitzer} Space Telescope \citep{Werner04} along with two photometers,  Infrared Array Camera (IRAC; \citealt{Fazio04}) and Multiband Imaging Photometer for \textit{Spitzer} (MIPS; \citealt{Rieke04}). The IRS performed more than $20\,000$ observations over the cryogenic mission lifetime (December 2003 - May 2009), corresponding to $\sim15\,000$ distinct targets of various kinds (Tables\,\ref{tab:observations}, \ref{tab:stats}). 
The IRS observed between $\approx5$ and $\approx38$\mic\ in two low-resolution modules ($R=\lambda/\Delta\lambda\sim60-120$) and $\approx10$ and $\approx37$\mic\ in two high-resolution modules ($R\sim600$). The main properties of these modules are described in Table\,\ref{tab:modules}. Most observations ($\approx85$\%) were performed in staring mode, i.e., as single sources or groups (``clusters'') of individual sources. The remaining corresponds to spectral mappings. 

\begin{table}
\begin{center}
  \caption{Spitzer/IRS observations. }
  \label{tab:observations}
  \begin{tabular}{lll}
  \hline
  \hline
              &  AORkeys & Objects$^{\rm a}$ \\
  \hline
 High-res & 7192 / 8419  & 4219 / 5075  \\
 Low-res & 13565 / 16040 & 10308 / 12129  \\
 \hline
 Total$^{\rm b}$ & 17850 / 21337 & 12390 / 14582 \\
  \hline
  \end{tabular}
  \tablecomments{For each entry we provide the number of observations \\
  performed in staring mode and the total (staring and mapping). }
\tablenotetext{a}{Object names as given by the observer. } 
\tablenotetext{b}{Some AORkeys were observed in both high- and low-resolution.  } 
\end{center}
\end{table}

The Cornell Atlas of Spitzer/IRS Sources (CASSIS; \url{http://cassis.sirtf.com}), presented in \cite{Lebouteiller11a} (hereafter L11), provides users with every low-resolution spectra observed by the IRS in staring mode. The pipeline performs automated decisions concerning the background subtraction and the choice of extraction method best adapted to the source spatial extent. For unresolved sources, the optimal extraction scales the point-spread function (PSF) to the data spatial profile and provides the best signal-to-noise ratio (S/N) as compared to the full aperture extraction, with an improvement by a factor of two for sources $\lesssim300$\,mJy. 
Furthermore, thanks to the super-sampled PSF, it became possible to perform optimal extraction for any source position along the slit. The super-sampled PSF also allows users to investigate complex source configurations (blended sources with/without extended emission component, sources shifted in the dispersion direction). While CASSIS provides the integrated spectra in such complex cases, the Spectroscopic Modeling Analysis and Reduction Tool (SMART; \citealt{Higdon04,Lebouteiller10}) can be used for a highly-customizable manual extraction allowing source disentanglement. 

CASSIS represents a tool of important legacy value for preparing and complementing observations by future IR telescopes, in particular the \textit{James Webb} Space Telescope (JWST). The online CASSIS database allows users to download spectra of publishable quality, and a local access to the full database is offered on request for large datasets. Since publication of L11, several updates have been made to the low-resolution pipeline (see Appendix\,\ref{app:lowres}). CASSIS has been used extensively for massive dataset analysis or specific targets (e.g.,  \citealt{Gonzalez15,Brisbin14,Lyu14,Sargsyan14,Sargsyan12,Brown13,Farrah13,Feltre13,Gonzalez13,Hernan12,Hurley12,LeFLoch12,Weedman12}).

\begin{table*}
\begin{center}
  \caption{Number of observations and distinct sources in the CASSIS atlas per scientific category.}
  \label{tab:stats}
  \begin{tabular}{l ll l}
  \hline
  \hline
 Category & Low-res & High-res & Total \\
  \hline
 ``Cosmology'' & 3 / 3   &  0 / 0 & 3 / 3   \\
  ``Cosmic infrared'' & 36 / 36 & 0 / 0 & 36 / 36 \\
 ``Galaxy clusters'' & 49 / 61    &  11 / 12 &  55 / 73   \\
 \hline
   ``High-z galaxies'' & 974 / 1258   &  35 / 40 & 993 / 1298  \\
 ``Intermediate-z galaxies'' & 838 / 844   & 45 / 46  & 862 / 890   \\
 ``Nearby galaxies'' &  462 / 488  & 207 / 216 &  603 / 704   \\
  ``Local Group galaxies'' & 536 / 561   &  26 / 27 & 542 / 588  \\
   ``Galactic structures'' & 13 / 13   &  0 / 0  & 13 / 13 \\
  \hline
 ``Interacting, mergers'' & 159 / 164   &  96 / 97  & 201 / 261 \\
  ``AGN, quasars, radio-galaxies'' & 1394 / 1559 &  490 / 538 & 1549 / 2097  \\
 ``ULIRGS, LIRGS'' &  611 / 632  & 429 / 440 &  777 / 1072  \\
 ``Starburst galaxies'' & 128 / 133   &  32 / 34 & 153 / 167  \\
 \hline
 ``Extragalactic jets'' & 1 / 5   & 0 / 0 & 1 / 5   \\
  ``Gamma-ray bursts'' & 4 / 4  &  2 / 2   & 4 / 6 \\
  ``Compact objects'' & 46 / 59   &  10 / 11  & 49 / 70 \\
  ``ISM            & 830 / 893   & 279 / 290  &  941 / 1183  \\
  ``H\2\ regions'' & 55 / 59    & 12 / 15  & 56 / 74  \\
\hline    
 ``Extragalactic stars'' & 164 / 165 &  6 / 6 & 170 / 171 \\
 ``Stellar population'' & 161 / 218    &  6 / 6  & 167 / 224 \\
``Massive stars'' & 152 / 159   &  79 / 81 & 198 / 240  \\
``Evolved stars'' & 1129 / 1304   &   893 / 1166 & 1581 / 2470  \\
``Brown dwarfs'' &  250 / 290    & 27 / 28 & 254 / 318  \\
\hline
 ``Star formation'' & 527 / 582   &  95 / 102 & 590 / 684  \\
 ``Young stellar objects'' & 1532 / 1691   &  336 / 352 & 1655 / 2043  \\
 ``Circumstellar disks'' &  2378 / 2524   & 477 / 531  &  2616 / 3055 \\
 \hline
 ``Extra-solar planets'' & 2 / 7    &  0 / 0 & 2 / 7  \\
 ``Planets'' & 24 / 32  & 16 / 18  & 24 / 50 \\ 
 ``Satellites'' & 22 / 23  &  14 / 15  & 24 / 38  \\
 ``Asteroids'' & 166 / 170  & 0 / 0 & 166 / 170 \\
 ``Kuiper belt objects'' & 31 / 31  & 0 / 0 & 31 / 31 \\
 ``Near-earth objects'' & 12 / 12  & 0 / 0 & 12 / 12 \\
 ``Comets'' & 41 / 54  & 48 / 59 & 77 / 113 \\
  \hline
  Total & 12720 / 14034   & 3623 / 4132 & 14405 / 18166    \\
  \hline
  \end{tabular}
  \tablecomments{Here we consider observations performed on sources within groups (``cluster mode observations'') as distinct observations, since several observations can be part of a single AOR. For this reason, the number of observations in this Table differs from what is given in Table\,\ref{tab:observations}. 
  For each category, we provide the number of distinct sources (calculated by using a separation threshold of $>4\arcsec$ from other sources within the same category) and the total number of observations. The full list of programs with their assigned category can be found at \url{http://isc.astro.cornell.edu/Smart/ProgramIDs}. Since only one category was assigned to any given program, some sources may have a false category identification. }
\end{center}
\end{table*}

In the present paper, we describe optimal extraction for the IRS high-resolution observations using a newly determined empirical super-sampled PSF. 
The two high-resolution modules use echelle spectroscopy as opposed to long-slit spectroscopy for low-resolution. In the following we refer to ``aperture'' for \textit{short-high} (SH) and \textit{long-high} (LH) as opposed to ``slit'' for \textit{short-low} (SL) and \textit{long-low} (LL). The high-resolution modules contain $10$ spectral orders (see Table\,\ref{tab:modules} and Figure\,\ref{fig:background_pattern} for a description of the detectors). Staring observations work the same way as for low-resolution observations, i.e., a source is observed in two nod positions, located at about $1/3$ and $2/3$ of the aperture length. 
With a spectral resolution $\sim10$ times higher than SL and LL, high-resolution observations are ideal for spectral line measurements and identification (and disentanglement) of narrow features that may be blended in the low-resolution spectra. For comparison, the FWHM in SH and LH observations, $\approx350-500$\,\kms\ depending on the spectral order, is somewhat larger than \textit{Herschel}/PACS ($60-320$\kms). Furthermore, since higher spectral resolution effectively results in lower signal-to-noise on the continuum for a given exposure time, high-resolution observations targeted mostly nearby bright sources (Table\,\ref{tab:stats}). The differences between high- and low-resolution observations performed with the IRS translate into several important differences as compared to the pipeline for low-resolution data that was presented in L11.

\begin{table*}
\begin{center}
  \caption{Main properties of the Spitzer/IRS modules. }
  \label{tab:modules}
  \begin{tabular}{cccccc}
  \hline
  \hline
  Module & $\lambda/\Delta\lambda$ & Aperture size  & Pixel scale  & Order(s) & $\lambda_{\rm min}-\lambda_{\rm max}$ \\
   &  & ($\arcsec$)  & ($\arcsec$) &  &  (\mic) \\
  \hline
  SL & 60-127 & 3.7$\times$57 & 1.8& 1 & 7.4 - 14.5  \\
   &  & & & 2 & 5.2 - 7.7   \\
   &  & & &  bonus & 7.3 - 8.7 \\
  \hline
  LL & 60-127 & 10.7$\times$168 & 5.1 & 1 & 19.5 - 38.0  \\
   && &   & 2  & 14.0 - 21.3  \\
   & & & & bonus  & 19.4 - 21.7   \\
  \hline
  SH & 600 & 4.7$\times$11.3 & 2.3& 11-20 & 9.9 - 19.6  \\
  \hline
  LH & 600 & 11.1$\times$22.3 & 4.5 & 11-20 & 18.7 - 37.2 \\
  \hline
  \end{tabular}
\tablecomments{More information can be found at \url{http://cassis.sirtf.com/atlas/irs_pocketguide.pdf}.}
\end{center}
\end{table*}

We first present the pipeline steps related to the detector images in Sect.\,\ref{sec:imsteps}. We then explain the full aperture extraction in Sect.\,\ref{sec:fullap} and optimal extraction in Sect.\,\ref{sec:optext}. In Section\,\ref{sec:opt_extended} we describe how the pipeline decides the best extraction method based on the source spatial extent. Finally, the post-processing steps at the spectrum level are described in Sect.\,\ref{sec:specsteps}.

\section{Detector image processing}\label{sec:imsteps}

\subsection{Individual exposures}\label{sec:discard}

The CASSIS pipeline uses the Basic Calibrated Data (BCD) images as starting products\footnote{Tests were made with ``droop'' images (identical to BCD images except lacking the flat-field), but there were no visible improvements in the final products.}, along with the corresponding uncertainty images and the bad pixel mask. The BCD images are produced by the \textit{Spitzer} Science Center BCD pipeline. We refer to L11 for details on BCD images. 

Individual exposure times for SH are $6$, $30$, $120$, or $480$ seconds. Exposure times for LH are $6$, $14$, $60$, or $240$ seconds. 
There is one set of data/uncertainty/mask images per exposure. The mask reflects possible problems identified by the pipeline. Before the exposures are combined (Sect.\,\ref{sec:ip_expcomb}), the pixel masks are first compared over the exposures. Some pixels may have a lower mask value in some exposures, and therefore indicate more reliable values. For each pixel of the detector image, we select only the exposures having the mask values $<256$. Pixels with higher mask  value (corresponding to non-correctable saturation, missing data in downlink, one or no usable ramp planes, or pixels for which the stray-light removal or cross-talk correction could not be performed) are ignored in the other exposures. For the vast majority of cases, the lowest mask value is null (i.e., the pixel flux is reliable).

\subsection{Exposure combination}\label{sec:ip_expcomb}

For a given module, order, and nod position, the flux of each detector pixel is compared over the exposures. For each pixel, the exposures in which the mask value is relatively higher are  discarded (Sect.\,\ref{sec:discard}), thereby selecting only the most reliable exposures for the combination. The final pixel flux value is the weighted-mean over the selected exposures. Weights are given according to the sequence number of the exposure because the first exposures are most affected by patterns and gradients present in the detector background (Sect.\,\ref{sec:detbg}). The relative weights are $1$, $3$, $5$ for the three first exposures, and $6$ for the remaining ones. The same weights are applied when the number of exposures is small (e.g., weights $1$ and $3$ for two exposures).

Since uncertainties on the individual pixels in exposure images may have been underestimated by the BCD pipeline, a simple quadratic sum of the uncertainties is not always accurate. For this reason, we calculate the uncertainties on the combined image as the maximum between the standard deviation of the mean of the pixel values over the exposures and the quadratic sum of the uncertainties.

\subsection{Detector background and order corrugation}\label{sec:detbg}

\begin{figure}[b!]
\includegraphics*[angle=0,width=8cm]{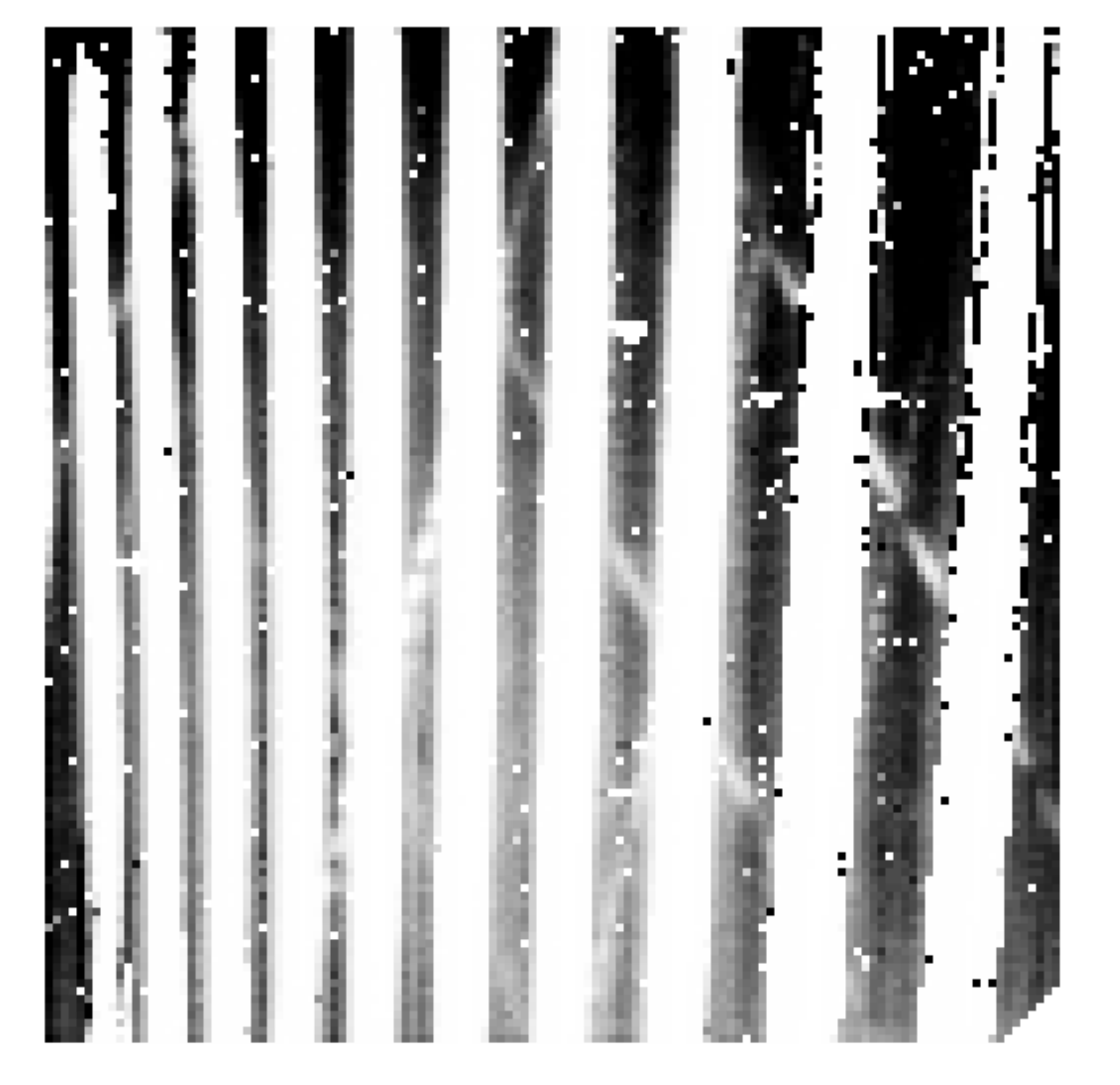}
\caption{LH detector image ($128\times128$\,px$^2$) of a heavily saturated source. Vertical white bands show the $10$ individual spectral orders in which the source is observed. For each spectral order, the wavelength axis is approximately vertical and the cross-dispersion axis is approximately horizontal. In the text we refer to rows and columns to describe the detector rows and columns in a given spectral order. 
The detector shows in this case artifacts in the form of several diagnonal streaks. This artifact is different from the background gradient presented in Fig.\,\ref{fig:background}. }
\label{fig:background_pattern}
\end{figure}

\begin{figure*}
\begin{centering}
\includegraphics*[angle=0,width=14cm,height=3.5cm,trim=0 0 0 0]{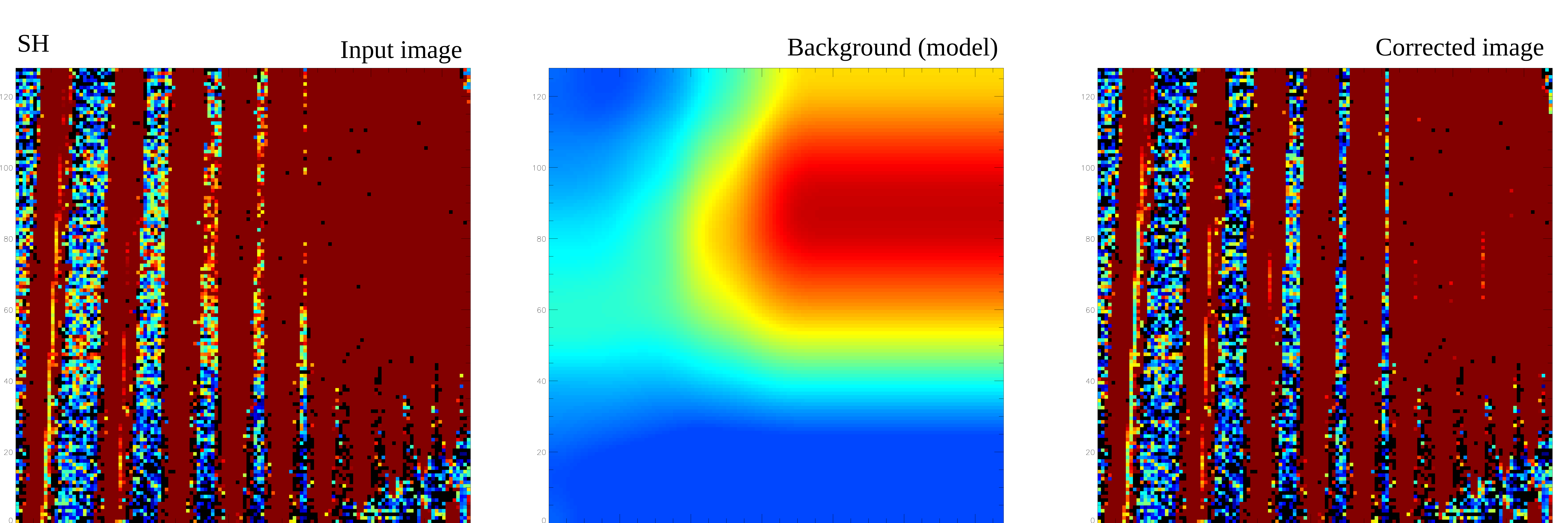}
\includegraphics*[angle=0,width=14cm,height=3.5cm,trim=0 0 0 0]{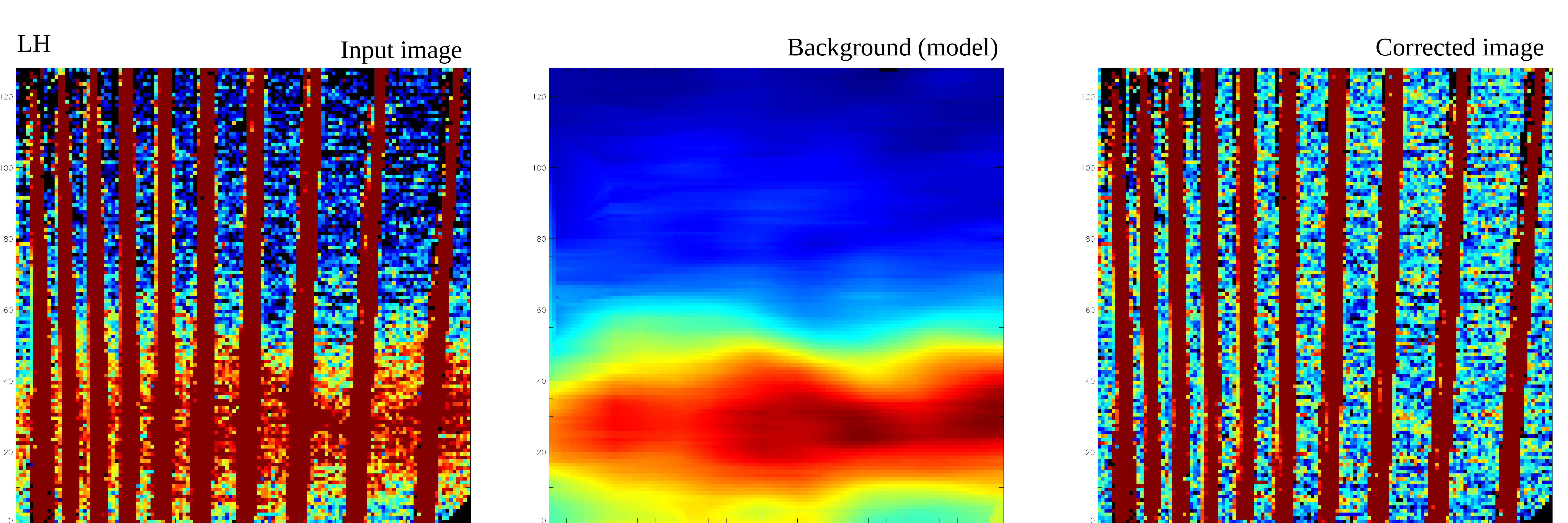}
\caption{Detector background removal for SH ($\xi$Dra, top) and LH (HD\,97300, bottom). \textit{Left} $-$ Input image with the contrast adjusted to visualize the low-level background gradient, \textit{center} $-$ background calculated with a 2D surface interpolation, \textit{right} $-$ corrected image. The scale is identical for all images in a given module (set to $20$\% of the median flux within the spectral orders). The SH spectral orders are not well separated, making it difficult to determine the background throughout the detector. As can be seen in the top-center image, one solution is simply to extrapolate the background from the left side of the detector. However, considering the uncertainties on the SH background and the fact that the SH gradient is less problematic than the LH one, we have chosen not to correct for it (see text).  }
\label{fig:background}
\end{centering}
\end{figure*}

The high-resolution detectors (in particular LH) sometimes show light traces across or gradients. These artifacts manifest themselves in the exposure image in two different ways:
\begin{itemize}
\item The background ``gradient'' is an unevenly distributed excess dark current throughout the detector (Fig.\,\ref{fig:background}). It affects as much as $\sim10$\%\ of the observations and is likely caused by a dark current residual. The detector background is generally more prominent during the observation of bright sources, with a fraction of the source's light scattered within the instrument. The detector may also be affected through latency by the prior observation of another, bright source. The intensity of the background gradient decreases systematically with the exposure number.
\item The background ``pattern'', made of diagonal streaks always appearing at the same locations (Fig.\,\ref{fig:background_pattern}) arises during or after the observation of extremely bright sources heavily saturating the detector. This pattern is mostly visible in a handful of observations and consists of several clumps/traces of pixels with a non-zero level. The pattern appears to be related to a permanent bias in some pixels of the detector. 
\end{itemize}
\noindent The background pattern, when visible, is associated with heavily saturated sources for which most, if not all, pixels within the spectral orders are masked out, making it an irrelevant artifact to fix. In the following, we describe possible corrections to the background gradient, which is the main cause of artifacts for the high-resolution observations, giving rise in particular to spectral order tilts if not corrected.

Removing a dedicated offset background image can mitigate the detector background artefact, but since the background gradient seems to depend on the source brightness, the correction is usually not satisfactory. Furthermore, dedicated offset background images are not always available or usable (Sect.\,\ref{sec:bgsub}) The difference between the two nod images (see Sect.\,\ref{sec:optext_flavors}) somewhat improves the removal of the background gradient but because of the separation in time between the two nod observations, a residual gradient still remains. 

In order to correct for the detector background gradient, one possibility is to use the \texttt{dark\_settle}\footnote{\url{http://irsa.ipac.caltech.edu/data/SPITZER/docs/dataanalysistools/darksettle/downloaddarksettle/}} algorithm provided by the SSC. This algorithm works for the LH detector only and computes a robust inter-order mean for a given row and smooths along the column. It then subtracts this mean from all the data in the row. In this way, the inter-order region for each row is set to zero. The \texttt{dark\_settle} algorithm partly removes  the detector background but residual large-scale variations are often still observed. Furthermore, the detector background gradient is not necessarily a simple slope along the columns. Therefore, we decided to implement a custom algorithm that computes the smoothed 2D surface of the detector using only the inter-order data, and interpolates over the spectral orders (Fig.\,\ref{fig:background}). In practice, the spectral orders are first masked using a conservative mask that ensures that no emission from the source is accounted for. This mask was created specifically for this purpose. The background image is then smoothed and interpolated over the spectral orders by means of a smoothed quintic surface. Note that for both \texttt{dark\_settle} and our own algorithm, the calculated correction has to be performed on the unflatfielded image. The interpolated surface calculated this way is not reliable for the SH detector since the spectral orders in this module are too close to each other on the detector image, as can be seen in Fig.\,\ref{fig:background}. Therefore we never attempt to correct for the detector gradient for SH.

Despite the success of mitigating the detector background gradient and significantly improving the LH spectra quality, our tests show that if the number of exposures is large enough, the combination of exposure images mitigates even better the gradient. The first exposure (in each nod observation) is indeed always the most affected by the gradient, which usually becomes negligible after $\gtrsim4$ exposures. For this reason, we chose to apply our background removal algorithm to LH data only when the number of exposures is $\leq2$. For a larger number of exposures, we simply rely on the exposure combination with relatively smaller weight given to the first exposures (Sect.\,\ref{sec:ip_expcomb}). The detector background is never removed for the SH detector, and the first exposures are simply given less weights. 

Despite the corrections applied above, some residual emission may remain that \textit{appears} as extended emission component in the aperture. If such a component is present, it is possible to remove it at a later stage when the optimal extraction is performed (Sect.\,\ref{sec:optext}).

\begin{figure*}
\hspace{-0.5cm}
\includegraphics*[angle=0,width=14cm]{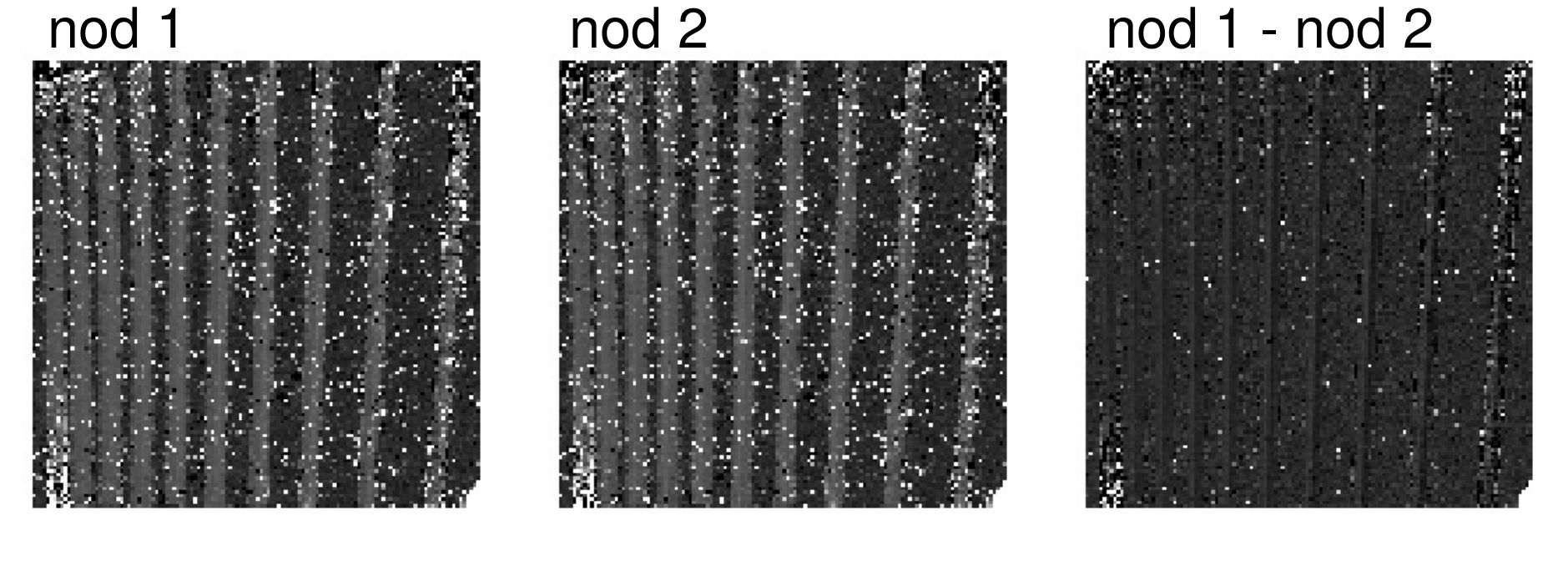}
\caption{Illustration of low-level rogue pixels being removed by differencing the two nod images. The scale is the same for all images. The resulting difference image can be optimally extracted if a differential PSF profile is used (Sect.\,\ref{sec:optext_flavors}). 
}
\label{fig:imdiff3}
\end{figure*}

\section{Full-aperture extraction}\label{sec:fullap}

We describe in this section how images are used to perform a full aperture extraction. This extraction method simply co-adds the pixels in a given pseudo-rectangle (area in the detector corresponding to one wavelength value) to compute the flux. Since the flux is integrated, the presence of bad pixels anywhere within the pseudo-rectangle is particularly harmful. Therefore, bad pixels need to be  identified and replaced. The cleaning is performed on the combined image of all exposures since bad pixels are replaced using neighbors whose flux is more reliable when exposures have been combined. 

We use the IRSCLEAN\footnote{\url{http://irsa.ipac.caltech.edu/data/SPITZER/docs/dataanalysistools/tools/irsclean/}} tool to substitute bad pixels in the following order:
\begin{itemize}
\item pixels with no values (NaNs) that may remain after exposure combination,
\item pixels with a high bad pixel mask value ($>256$; see Sect.\,\ref{sec:discard} for the description of the corresponding instrumental artifacts),
\item bad pixels and rogue pixels (i.e., pixels with a significant variations in their responsivity over time) flagged in the campaign mask,
\item pixels with a large uncertainty ($>10$ times the median uncertainty in the image),
\item negative pixels, if $>10$ times the median uncertainty. 
\end{itemize}

\noindent The new pixel value is calculated by the IRSCLEAN algorithm mainly based on neighboring pixels, although in some cases IRSCLEAN cannot substitute every eligible pixel due to clustering. Uncertainties and mask values are propagated for each step. The full-aperture extraction is performed using the standard tool in SMART. Contrary to optimal extraction (Sect.\,\ref{sec:optext}), the flux determination in the full-aperture extraction does not depend on the individual pixel uncertainties since the pixel values are simply summed. An error on the flux is ultimately calculated using the quadratic sum of the pixel uncertainties in the pseudo-rectangle.

There is no background subtraction by default for full-aperture extraction. Although dedicated offset background observations may exist, they are not considered in the current version of CASSIS (Sect\,\ref{sec:bgsub}) and the extracted spectrum simply corresponds to the addition of the source spectrum and any background emission that may be present. Note that it is always possible to download the full-aperture extracted spectrum of the dedicated background observation (if known) separately and subtract it from the science source spectrum. In this case it is preferable to subtract the background at the image level to remove potentially bad pixels but the subtraction of the two spectra corrects for any emission not related to the nominal source.

\section{Optimal extraction}\label{sec:optext}

Optimal extraction uses the PSF profile to compare to the data spatial profile in order to calculate the flux density (see e.g., \citealt{Horne86}). Optimal extraction provides a spectrum with a higher signal-to-noise ratio when the source is unresolved. In the following, we describe how bad pixels are handled by the algorithm, how the super-sampled PSF is created, and how optimal extraction is performed on the data. 

\subsection{Bad pixels}

For the optimal extraction, and contrary to full aperture extraction (Sect.\,\ref{sec:fullap}), the bad pixels that were identified do not have to be substituted since the PSF is fitted to the spatial profile of the object at any  wavelength. Therefore, gaps in the spatial profile are not problematic as long as the spatial profile is sufficiently sampled. In cases when the spatial profile cannot be reliably analyzed because too many pixels are missing, the corresponding wavelength row is flagged as being unusable during the extraction step (Sect.\,\ref{sec:optext_core}). Another difference with the treatment of bad pixels between full aperture and optimal extraction is that uncertainties on individual pixels are used to determine the flux for the optimal extraction. Therefore, bad pixels are identified using the same steps as for full aperture (Sect.\,\ref{sec:fullap}) \textit{except for the pixels with large uncertainties} which are kept as such for optimal extraction.

Many transient pixels, also referred to as low-level ``rogue'' pixels, are not flagged and are usually best corrected for by removing a background exposure. In the majority of observations, no dedicated background pointing was performed (Sect.\,\ref{sec:bgsub}). In such cases, it is still possible to subtract the other nod observation (Fig.\,\ref{fig:imdiff3}) as long as one accounts for the resulting differential spatial profile for extracting the flux (Sect.\,\ref{sec:optext_core}).

\begin{figure}[h]
\includegraphics*[angle=0,width=7.8cm,clip=true]{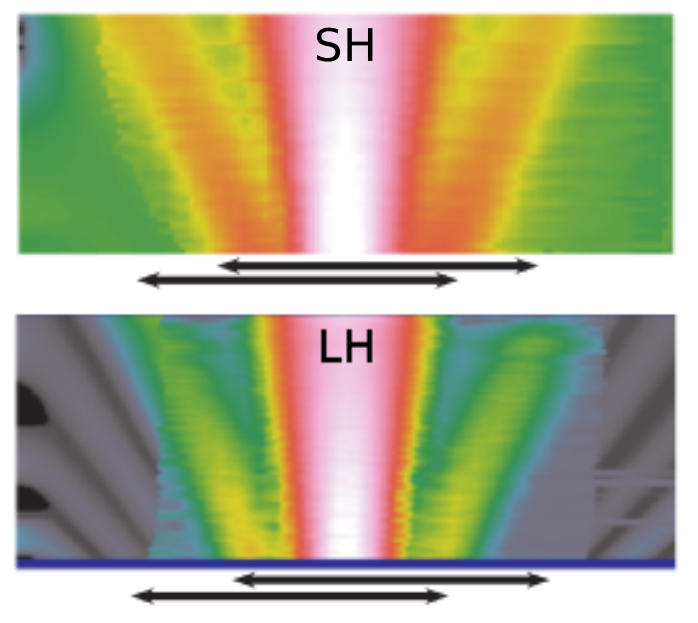}
\caption{The super-sampled PSF for SH and LH. Wavelength increases from bottom to top. The black lines below indicate the approximate size of the aperture and the relative location of the super-sampled PSF within the aperture at the two nod positions. The super-sampled PSF was created on a grid with sub-pixels three times smaller than the pixel in the SH and LH detectors. The data beyond the secondary peak is that of the model PSF (see text). Such data is not used in standard observations in which the source lies in the aperture.  }
\label{fig:sspsf}
\end{figure}

\subsection{Super-sampled PSF}\label{sec:sspsf}

A super-sampled PSF, either theoretical or empirical, is necessary for the optimal extraction of sources located anywhere in the aperture. The super-sampled PSF is built from mapping observations of point-like sources ($\xi$ Draconis for SH and Sirius for LH) scanned along and across the apertures, with a step size smaller than the size of a pixel. We  performed an iterative reconstruction of the high-resolution spatial profile from the under-sampled data. We refer to \cite{PinheirodaSilva06} and L10 for details on the algorithm. The resolution on the PSF was increased by a factor of three for SH and LH.

\begin{figure*}
\includegraphics*[angle=0,width=16.8cm]{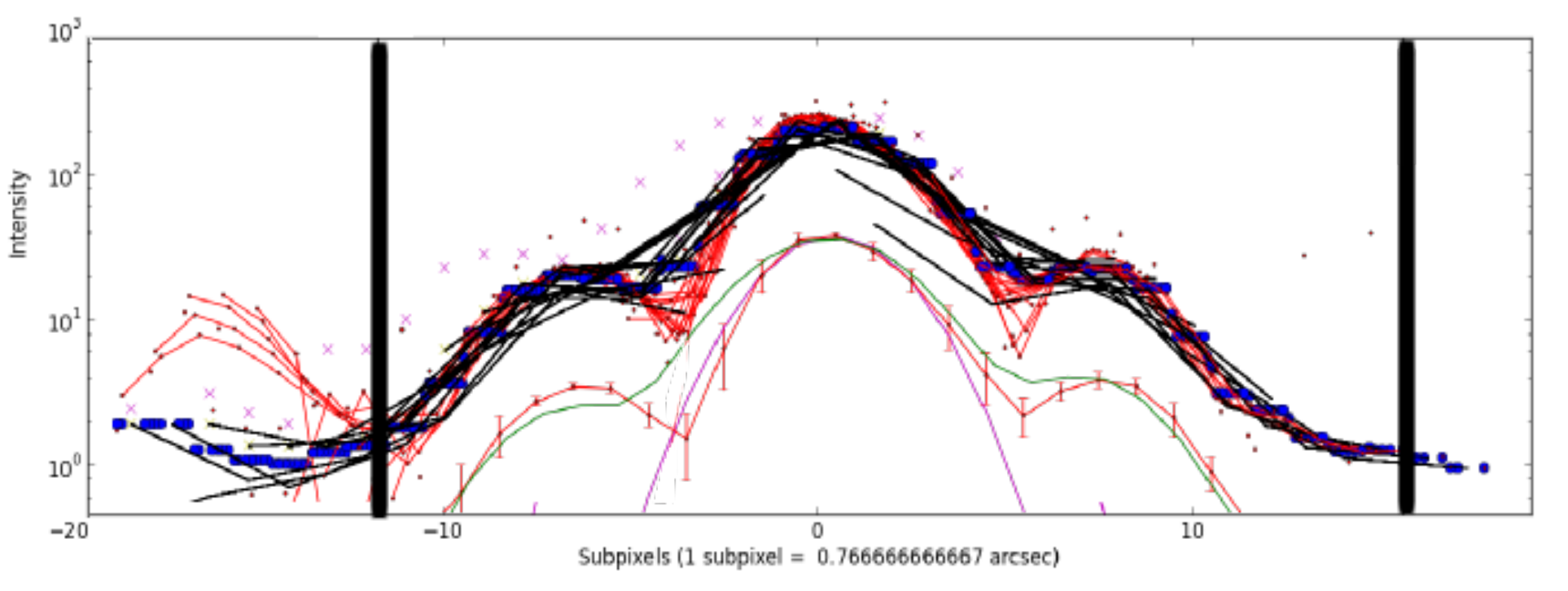}
\caption{Illustration of the super-sampled PSF creation steps for one row of one spectral order in the SH detector (pixels in one row of a given spectral order have approximately the same wavelength). The thick vertical lines mark the region in which the PSF was calculated (the ``output window''). Black segments represent input observations (i.e., sequence of exposures with the star located at various positions in the aperture). Blue dots show the co-addition of all input data (used for bad data replacement). Fractions of the PSF are calculated within ``sub-windows'' that are shown with the red dots (covering the full sub-window) and red lines (covering only the part of the sub-window that will contribute to the final PSF). Purple crosses indicate image data that was discarded based on our data replacement algorithm. 
The final super-sampled PSF is shown by the shifted red profile in the lower part of the plot, along with the \texttt{STINY\_TIM} 2D PSF (collapsed in the dispersion direction; green), and a Gaussian fit to the core (purple).
}
\label{fig:sspsf_profile}
\end{figure*}

A major difference with the algorithm used for the low-resolution PSF described in L11 is that the SH and LH apertures are relatively small and miss a fraction of the desired PSF profile in any given observation. Figure\,\ref{fig:sspsf} shows the relative location of the PSF within the aperture for the two nod positions. Since the PSF is never fully sampled, we built the super-sampled PSF ``piece by piece''. For this (1) we first cut our desired output window (that eventually contains the final super-sampled PSF) into many overlapping sub-windows with the same size as the aperture, (2) calculated super-sampled PSFs for each of these output windows separately\footnote{Only the central part of the sub-windows are actually used, in order to mitigate edge effects related to systematic uncertainties at the edge of the aperture.}, and (3) combined them together into a final PSF covering the desired range. We also added various steps of regularization and data filtering/replacement in order to improve the quality of the final PSF. An illustration of the super-sampled creation process is shown in Fig.\,\ref{fig:sspsf_profile}.
The super-sampled PSF provides an important improvement over the model created from the  \texttt{STINY\_TIM}\footnote{\url{http://irsa.ipac.caltech.edu/data/SPITZER/docs/dataanalysistools/tools/contributed/general}} 2D PSF collapsed in the dispersion direction (green profile in Fig.\,\ref{fig:sspsf_profile}). .

\begin{figure}
\includegraphics*[angle=0,width=8cm,height=4.2cm,trim=60 60 20 30]{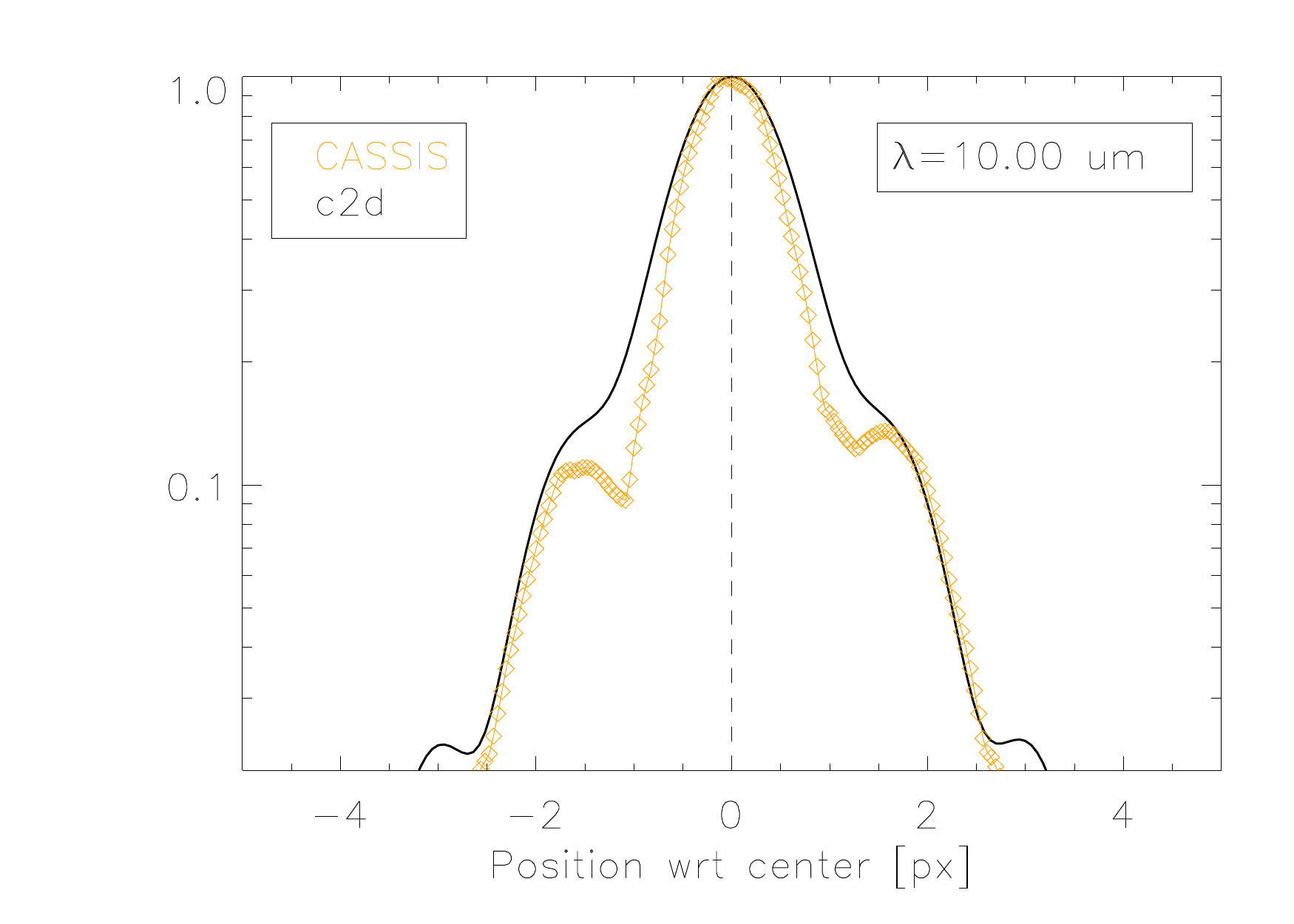}
\includegraphics*[angle=0,width=8cm,height=4.2cm,trim=60 60 20 30]{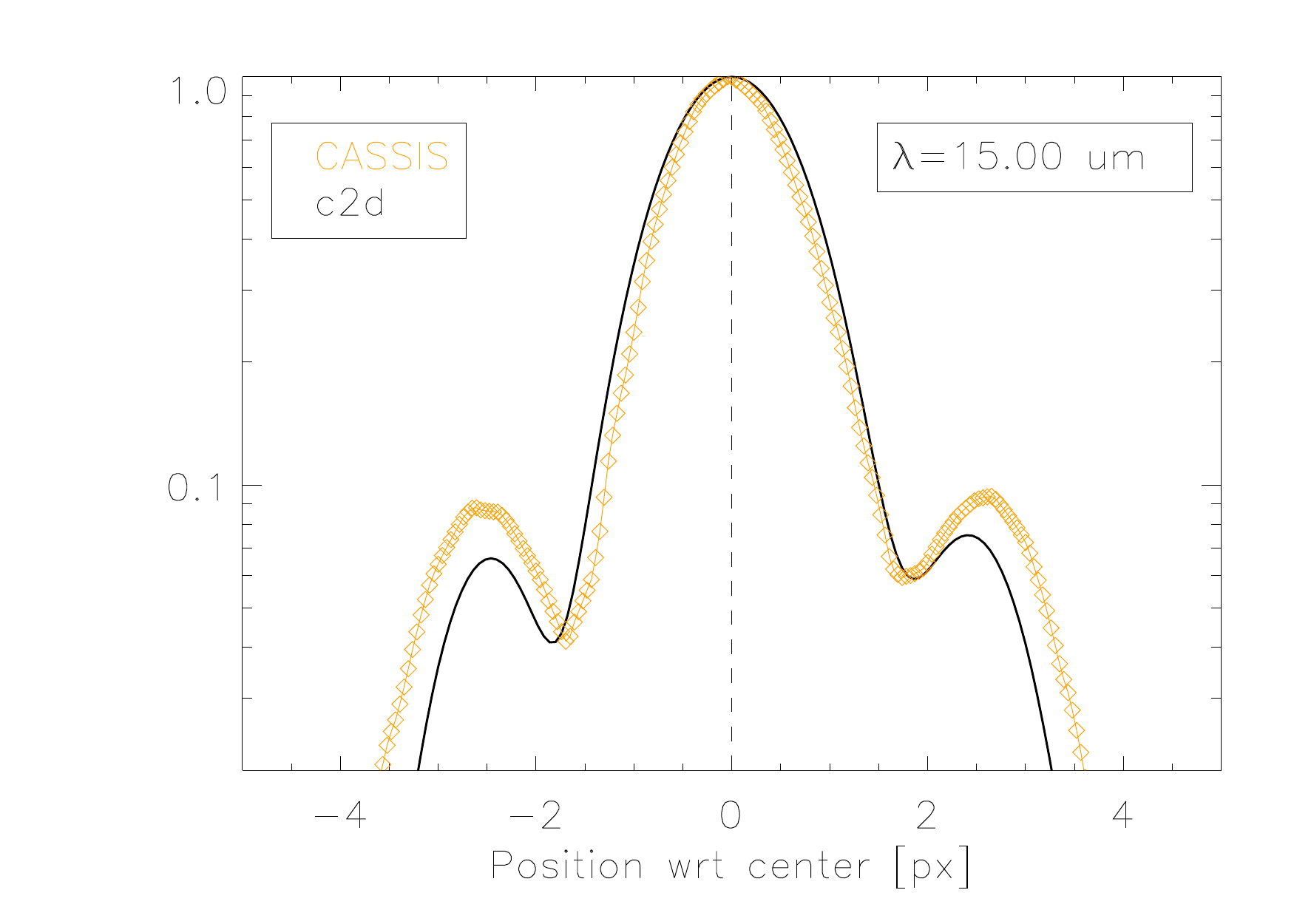}
\includegraphics*[angle=0,width=8cm,height=4.2cm,trim=60 60 20 30]{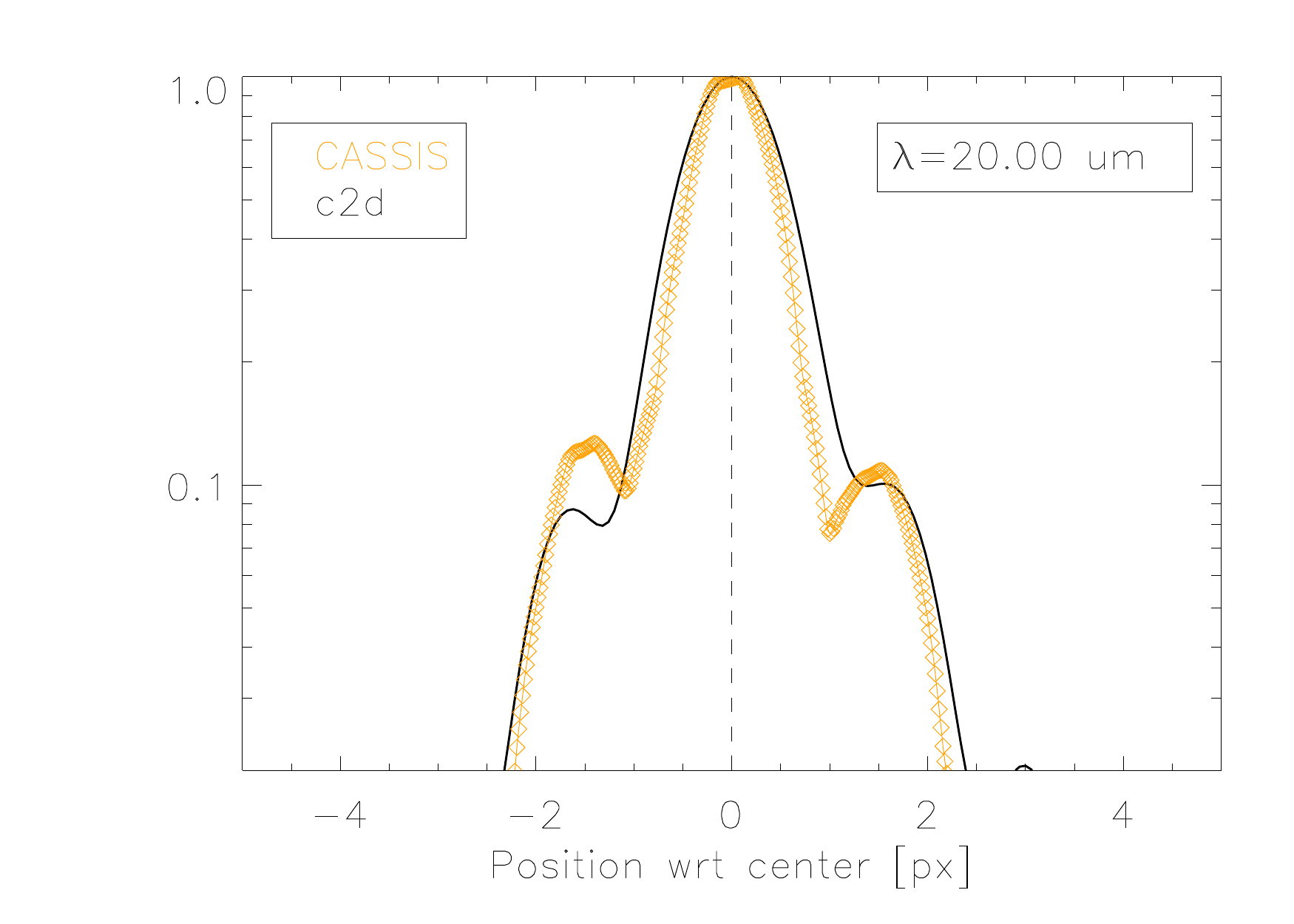}
\includegraphics*[angle=0,width=8cm,height=5.0cm,trim=60 0 20 30]{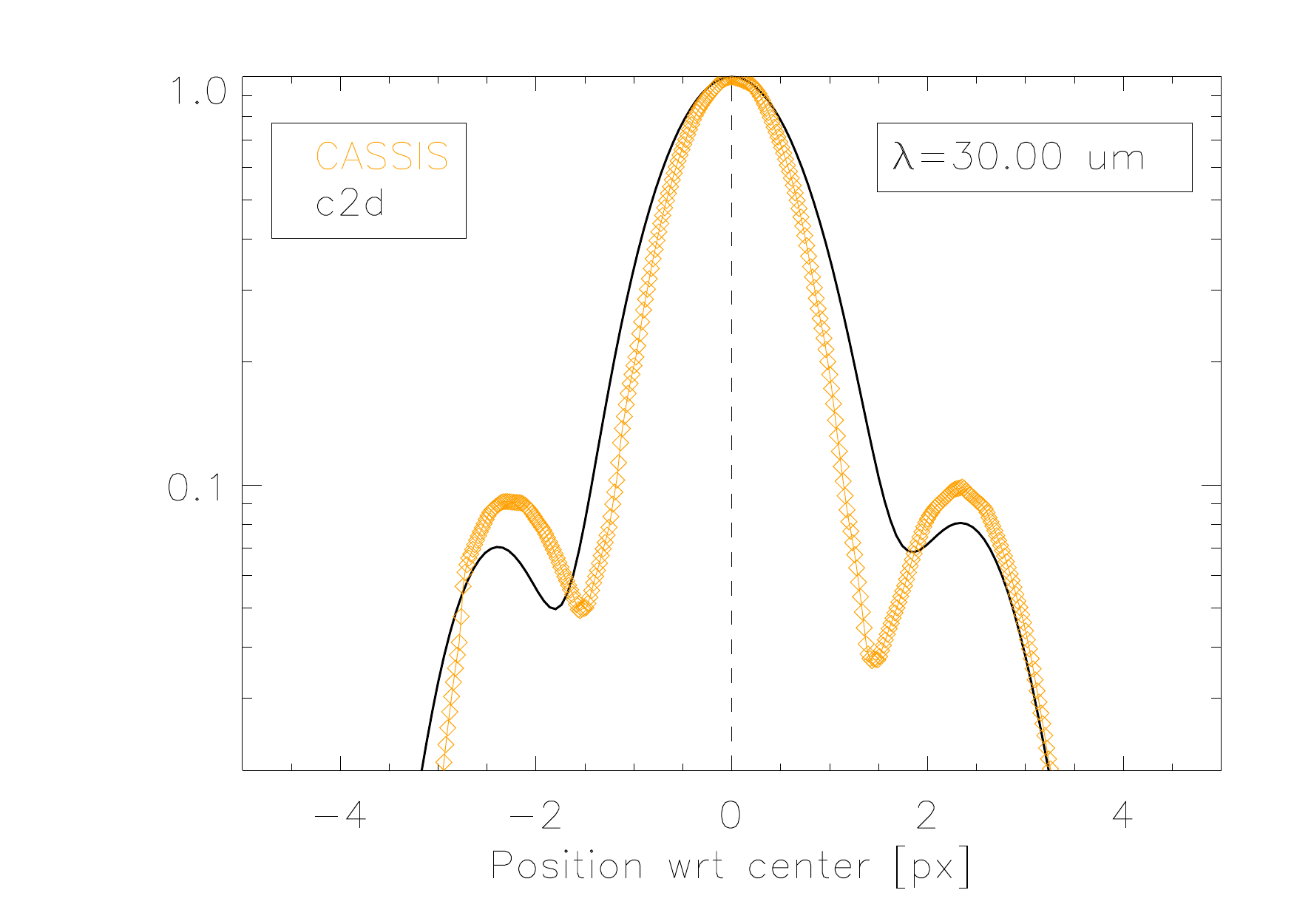}
\caption{Comparison between the CASSIS super-sampled PSF and the c2d analytical profile \citep{Lahuis07PhD} for selected wavelengths in SH ($15,20$\mic) and LH ($20,30$\mic). The CASSIS PSF reaches a higher resolution on the actual instrumental profile, resulting in the core being slightly narrower and the first Airy ring being slightly brighter. }
\label{fig:c2dcompa}
\end{figure}

Our approach to the PSF is different from the c2d project optimal extraction \citep{Lahuis07PhD,Lahuis07,Lahuis10}, the latter using an analytical cross-dispersion PSF (described as a cardinal sine function with a harmonic distortion component). A comparison of our super-sampled PSF to the c2d analytical instrumental profile reveals a slightly narrower PSF core and more power in the first Airy ring (Fig.\,\ref{fig:c2dcompa}). These differences result from the fact that the super-sampled PSF achieves a better resolution on the instrumental profile by improving on the originally low sampling of the PSF in the data, while the c2d PSF is calculated by fitting a model to the original (not super-sampled) data.

A super-sampled PSF was thus created for the first time for SH and LH that provides the profile of a point-source anywhere in the aperture (Fig.\,\ref{fig:sspsf}). While the sources are well-centered in the aperture in the \textit{dispersion} direction in most observations, we have also computed the super-sampled PSF at various positions across the \textit{dispersion} direction. Such profiles can be used to perform an optimal extraction of mispointed sources. For the CASSIS online repository, however, we assume the source to always be centered in the \textit{dispersion} direction. The source position along the \textit{cross-dispersion} direction is a free parameter (Sect.\,\ref{sec:sf}).

\subsection{Spatial under-sampling}\label{sec:undersampling}

The shortest wavelengths of SH and LH can be affected by spatial under-sampling. For these wavelengths, the FWHM of the PSF is on the order of one pixel, which requires the use of a super-sampled PSF (Sect.\,\ref{sec:sspsf}) together with an accurate source position. When a source shifts \textit{within} a given detector pixel, the intra-pixel responsivity can lead to significant variations in the estimated flux. Ignoring the under-sampling effect may result in wiggles in the extracted spectrum, which is the result of the spectral trace (position of the source centroid in the detector for a given spectral order) not being perfectly orthogonal with the detector axes.

For the low-resolution pipeline of CASSIS, the undersampling was apparent for LL2 (at wavelengths $\lambda\lesssim21$\mic), SL2 ($\lambda\lesssim8$\mic), and SL1 ($\lambda\lesssim14$\mic), by order of importance. The correction was performed by applying an empirical intra-pixel response function to the projected PSF on the detector grid, the corrective function being closer to unity with increasing wavelength. 

The smallest wavelength in SH is $\approx10$\mic, so we expect the under-sampling problem in SH to be equally important as for SL1, i.e., minor. The first release of the CASSIS high-resolution pipeline assumes that under-sampling effects in SH can be ignored. The smallest wavelength in LH is $\approx19$\mic, and under-sampling problems do appear for the shortest wavelengths ($\lesssim24$\mic), requiring the use of an empirical correction. The latter was performed the same way as for the low-resolution modules, with an intra-pixel responsivity decreasing with distance from the pixel center and with a correction decreasing with wavelength.

\subsection{Optimal extraction kernel}\label{sec:optext_core}

Our pipeline uses a super-sampled PSF (Sect.\,\ref{sec:sspsf}) and a multiple linear regression to fit the source spatial profile. The latter is reproduced by the combination of the super-sampled PSF itself and a large-scale emission that can be parametrized (which we choose as a polynomial of order $0$). The super-sampled PSF is first shifted to the source position, resampled, and finally scaled to the data profile. The scaling factor and large-scale emission are fitted simultaneously.

As for the low resolution algorithm in L10, the high-resolution algorithm uses a weighted multiple linear regression, but within an iterative process. The incomplete covering of the PSF profile and the higher occurrence of bad pixels in the high-resolution modules results in a relatively smaller number of available pixels as compared to the low-resolution modules. In the first iteration, the flux is calculated several times, by elevating the weight of every pixel in the core of the spatial profile. This gives up to $4-5$ flux determinations. Outlier flux values are then flagged, and the uncertainty on the corresponding pixel is increased, since it is likely bad. In the second iteration, its weight is thus reduced. The iterations continue until there are no more outliers. In practice, only one or two iterations are needed. 

\noindent Several complications may occur:
\begin{itemize}
\item The flux values determined using each pixel in a given row do not agree within errors. In that case, we give less weight to the pixel corresponding to the outlying flux value, and we perform another iteration. 
\item Pixels with no valid value are not considered for the fit. In order to reflect the uncertainty associated with missing information, we calculate the fraction of flux in the valid pixels compared to the expected flux. The uncertainty on the final flux determination is then scaled up by an empirical coefficient inversely proportional to this fraction.
\item The model profile is always positive but some pixels can have negative values. The model is therefore never able to accommodate the sign inconsistency, regardless of the weights. Hence, rather than letting the model be biased toward low flux values, we replace the bad pixel with a null value and we increase the error bar, if necessary, to accommodate with the old value.
\end{itemize}

\subsection{Source finder}\label{sec:sf}

The source position is first approximated from a Gaussian fit to the collapsed spatial profile over all the wavelengths of all spectral orders. A more accurate position is then found through an iterative process. The position is varied around this first guess, and for each position the super-sampled PSF is calculated (i.e., shifted at the right position and resampled). Residuals are then calculated the same way as for the low-resolution optimal extraction \citep{Lebouteiller10}. In short, the (collapsed) spatial profile of the image is compared to the (collapsed) spatial profile of the image minus the source model. The goal is to minimize the latter difference. The accuracy is typically less than $1/10$th of a pixel, i.e., similar to what is accomplished for the low-resolution algorithm.

\subsection{Extended background}\label{sec:optext_extended}

We refer to the extended backgound emission as the emission that uniformly fills the aperture and that, in most cases, is not associated to the nominal science source. 
The extended backgound emission within the aperture is either instrumental (e.g., residual from the detector background gradient; Sect.\,\ref{sec:detbg}) and therefore a function of the detector row index, and/or it is physical and therefore a function of wavelength. Extended background emission unrelated to the source can originate from zodiacal emission or high galactic latitude cirrus clouds. In some cases, extended emission may be physically associated with a point source, e.g., an active galactic nucleus and the host galaxy. 
Note that if the science source in the aperture is found to be extended, the full aperture extraction will be the default method and dedicated offset images need to be used to remove the unassociated background (Sect.\,\ref{sec:bgsub}). 

On first approximation, the extended background in the aperture takes the shape of a plateau underneath the PSF. The multiple linear regression algorithm allows for a constant term to be calculated simultaneously with the PSF scaling factor (Sect\,\ref{sec:optext_core}). For the high-resolution observations however, the number of degrees of freedom is relatively small and the spatial profile does not cover much of the extended background far from the PSF core. For this reason, a first iteration is performed in which the extended background level is derived using the multiple linear regression, and the derived background spectrum is then slightly smoothed as a function of the row index, before being removed from the spatial profile in a second iteration.

\begin{figure}
\includegraphics*[angle=0,width=8cm,height=5cm,trim=0 1cm 0 14.8cm]{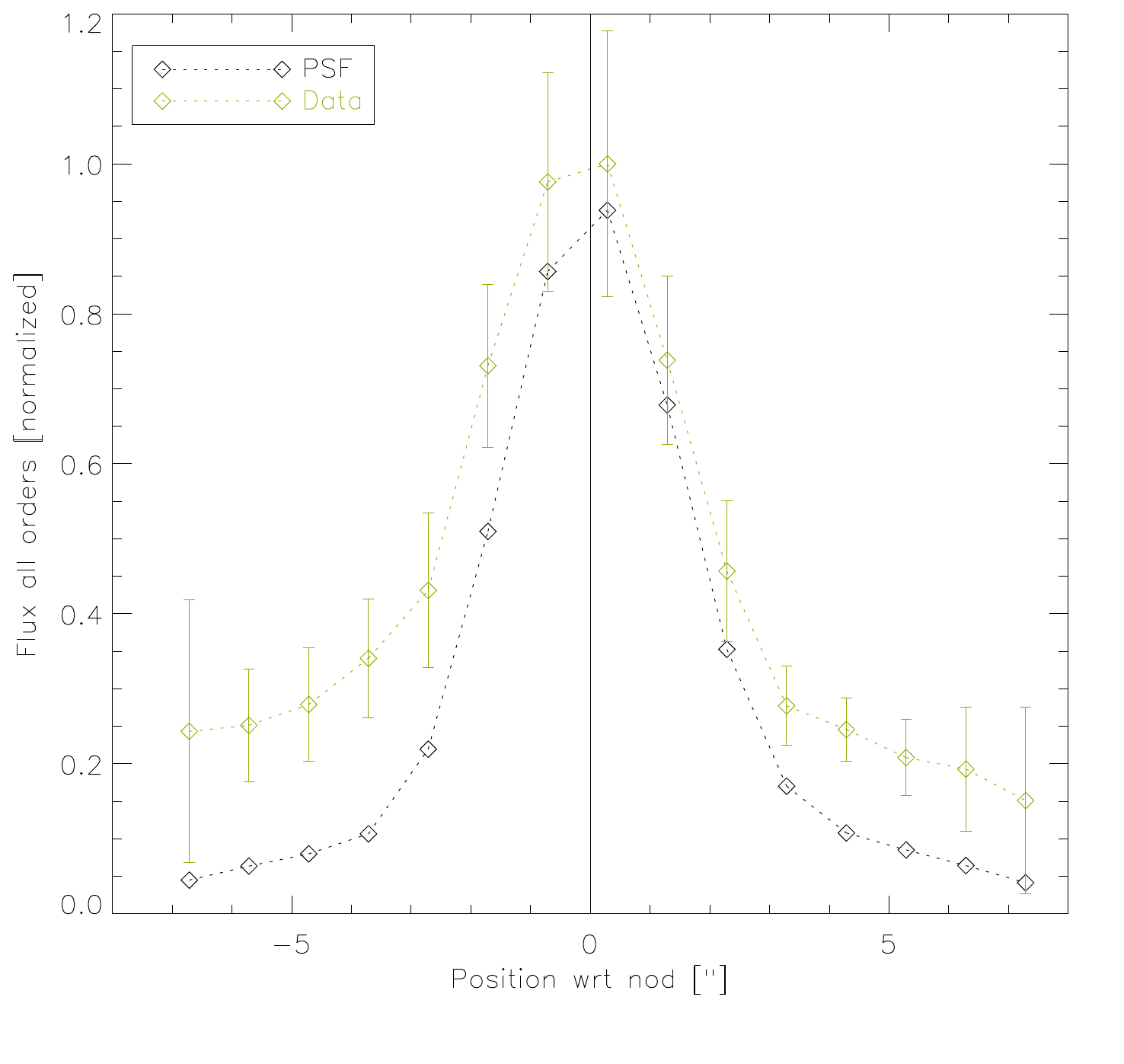}
\includegraphics*[angle=0,width=8cm,height=3.6cm,trim=0 1.4cm 0 0]{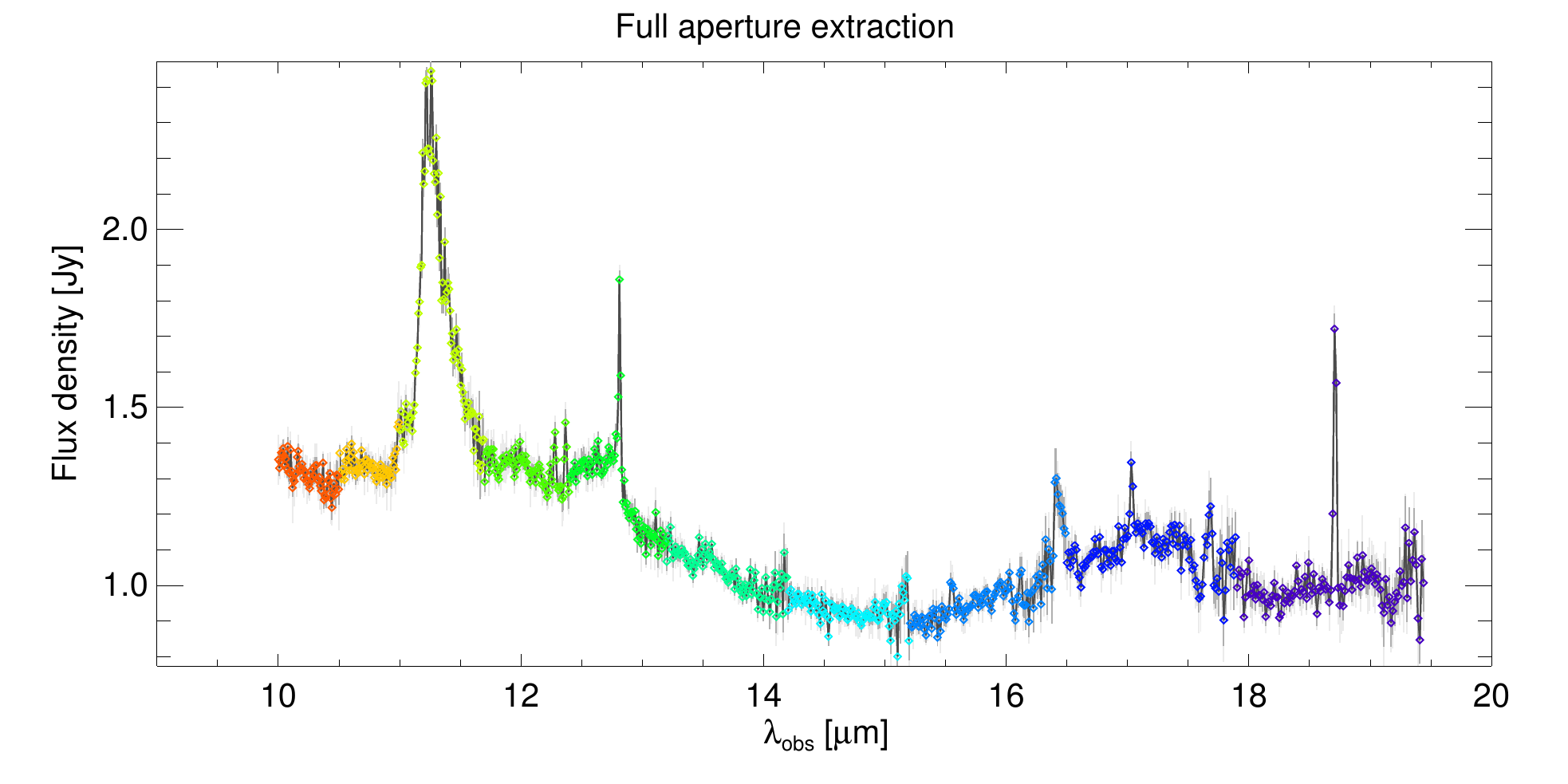}
\includegraphics*[angle=0,width=8cm,height=3.6cm,trim=0 1.4cm 0 0]{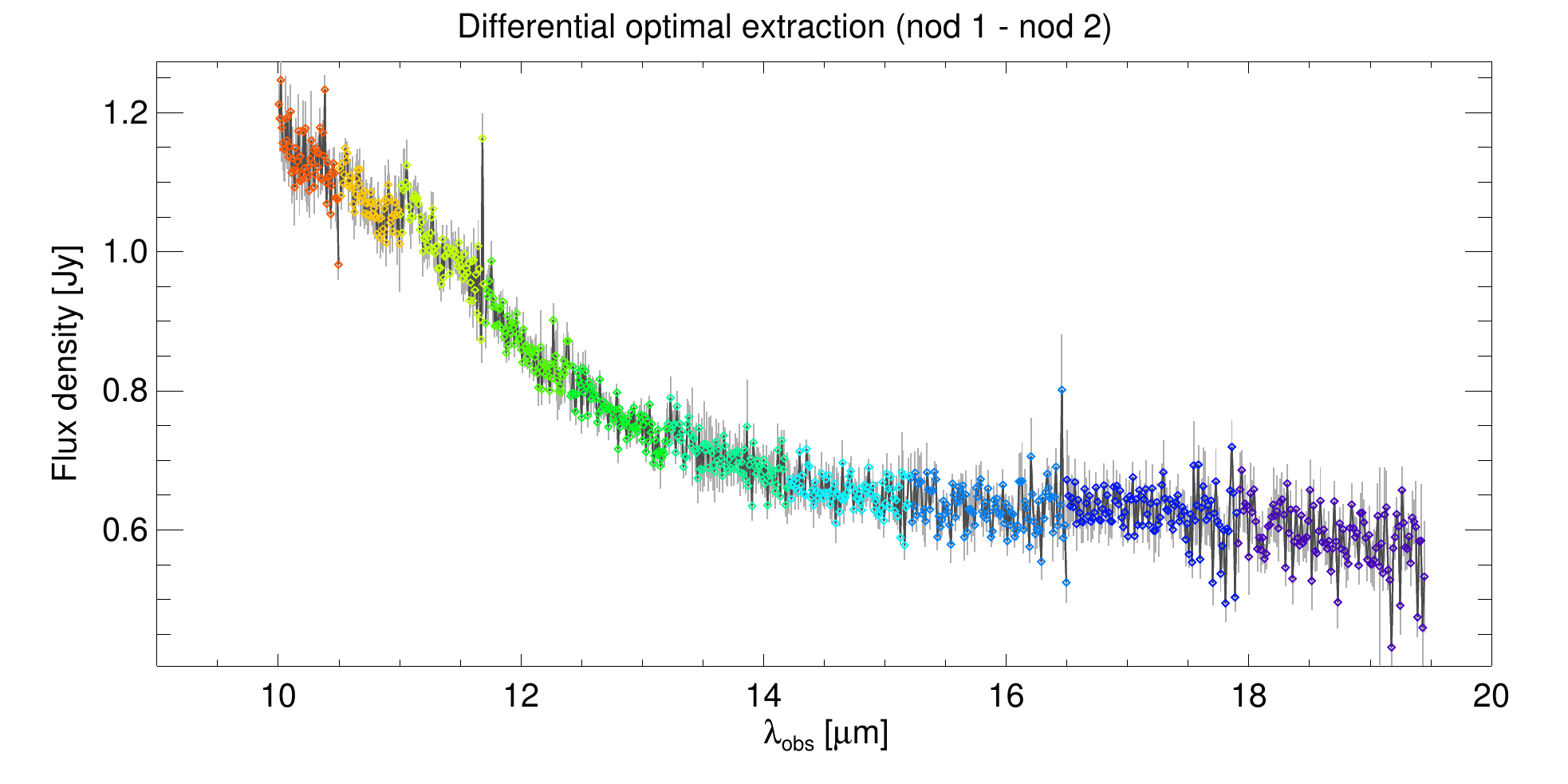}
\includegraphics*[angle=0,width=8cm,height=4.0cm,trim=0 0 0 0]{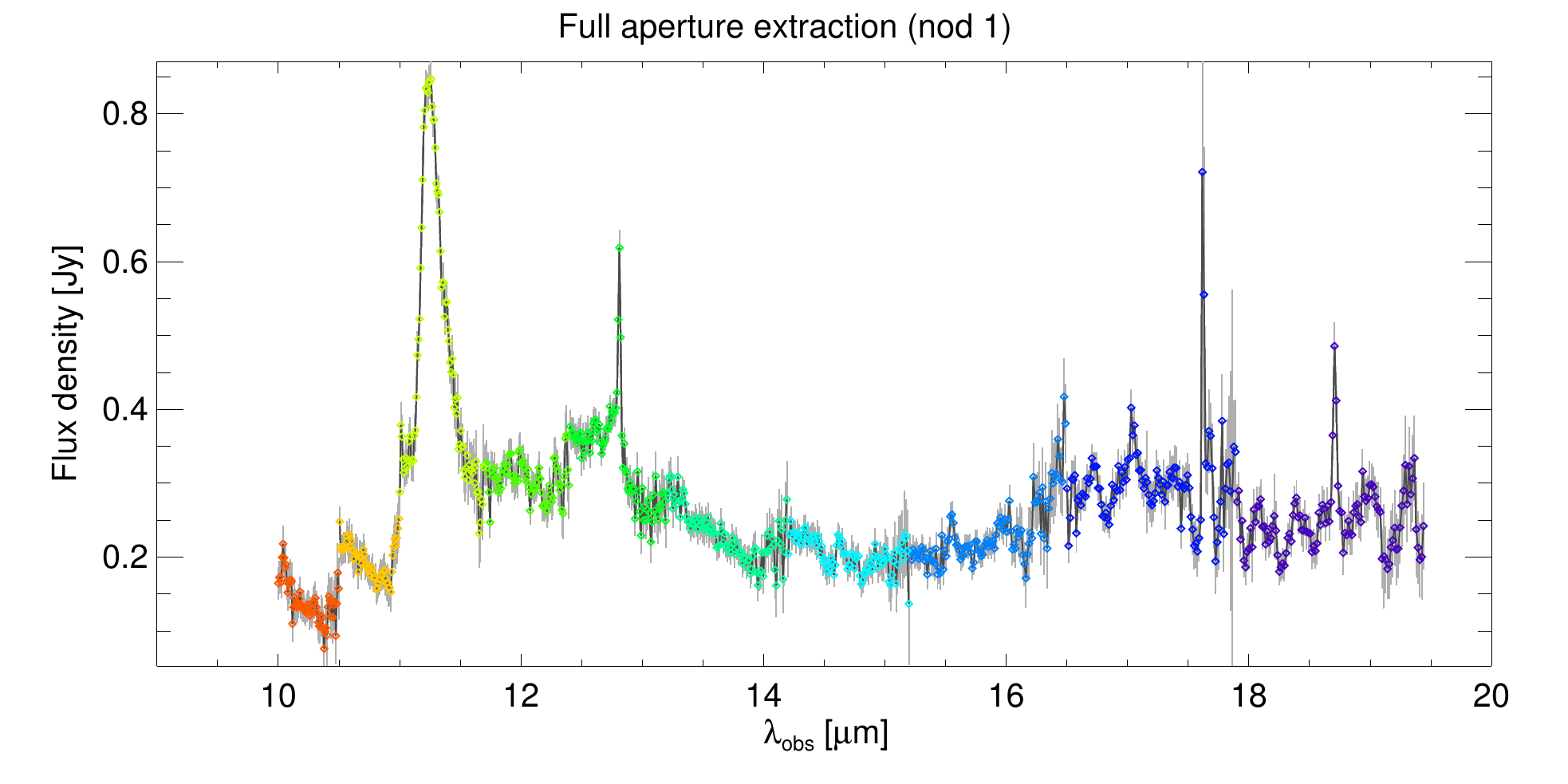}
\caption{The SH spatial profile of the Herbig Ae/Be star HD\,36917 (AORkey 11001600)  is shown on top; it consists of a point source and a background pedestal emission. The lower panels show from top to bottom the spectra from full aperture extraction, differential optimal extraction, and from the extended background. CASSIS optimal extraction extracts simultaneously the featureless point source and the extended background with bright emission from polycyclic aromatic hydrocarbons at $\approx11.3$\mic.  }
\label{fig:decomp}
\end{figure}

The philosophy of CASSIS is to provide the spectrum of point sources or extended sources. Apart from the removal of the extended background, no profile decomposition is available (e.g., multiple and blended point sources). 
In the website, users are encouraged to examine the spatial profile and validate the CASSIS approach. Furthermore, since in certain cases users may be interested in comparing the point source spectrum to that of the extended background (or investigating only the extended background), links are provided to download the extended background spectrum separately. Figure\,\ref{fig:decomp} illustrates how CASSIS disentangles the point-source emission from the extended \textit{physical} background in the observation of the star HD\,36917. The extended background is completely removed by CASSIS, resulting in the featureless point source spectrum and achieving the same result as the dedicated effort by \cite{Boersma08} for this particular source.

\subsection{Uncertainties on the flux}

For all flavors of optimal extraction (Sect.\,\ref{sec:optext_flavors}), the flux uncertainty for a given wavelength element is the error in the PSF scaling factor. This error is dominated by the uncertainties on the pixels of the image fed to the optimal extraction core. An uncertainty is also calculated for the extended background, since an error on the latter results in an error on the point-source flux. Other sources of uncertainties (fringe correction, nod combination, calibration) are also quantified by the pipeline.

\subsection{Optimal extraction methods}\label{sec:optext_flavors}

The optimal extraction as described in Sect.\,\ref{sec:optext_core} can be performed on various image products. The regular method consists in scaling the PSF directly to the data spatial profile (Sect.\,\ref{sec:reg}) while the differential method uses the subtraction between two nod images (Sect.\,\ref{sec:diff}). 

\subsubsection{Regular method}\label{sec:reg}

Optimal extraction can be performed on the two nod images \textit{individually}, producing two  independent spectra which can be merged eventually into a single spectrum. The main drawback of this method is that few pixels are available for scaling both the super-sampled PSF and the large-scale emission in the aperture simultaneously (see Sects.\,\ref{sec:optext} and \ref{sec:optext_extended}).

Optimal extraction can also be performed on the two nod images \textit{simultaneously}. In this case, we extract a single spectrum from the two images, taking advantage of a better sampling of the source spatial profile provided by the different nod positions in the detector. For this method, we assume that the PSF scaling factor and the background emission component will be identical for both nod images for a given wavelength in a given spectral order. The redundancy improves significantly the quality of the optimal extraction, in particular when bad pixels plague the image. The background emission is particularly better determined since by combining the two nod profiles, one can sample both sides of the PSF at the same time.  

The two methods described above are used in our pipeline in complementary ways, in the following order:
\begin{itemize}
\item Optimal extraction of the two nods \textit{individually} to find the source position in each nod observation (see explanations on source finder in Sect.\,\ref{sec:sf}). The background emission component is ignored for this step since it is not an important parameter for the source finder. 
\item Optimal extraction of the two nods \textit{simultaneously}, using the source positions from the previous step. This results in a single spectrum for the science object. The extended background spectrum is saved as an optional product.
\end{itemize}

\begin{figure}
\includegraphics*[angle=0,width=8cm,height=3.7cm,trim=0 15 0 0]{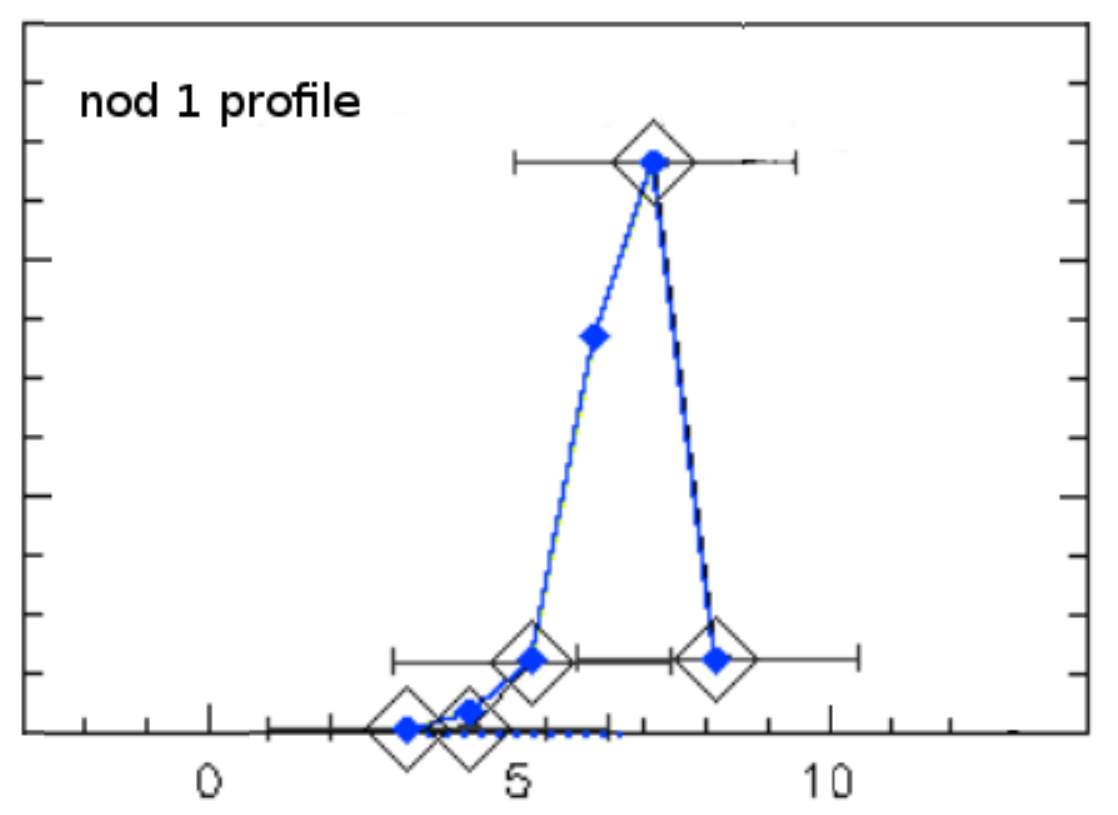}
\includegraphics*[angle=0,width=8cm,height=3.7cm,trim=0 15 0 0]{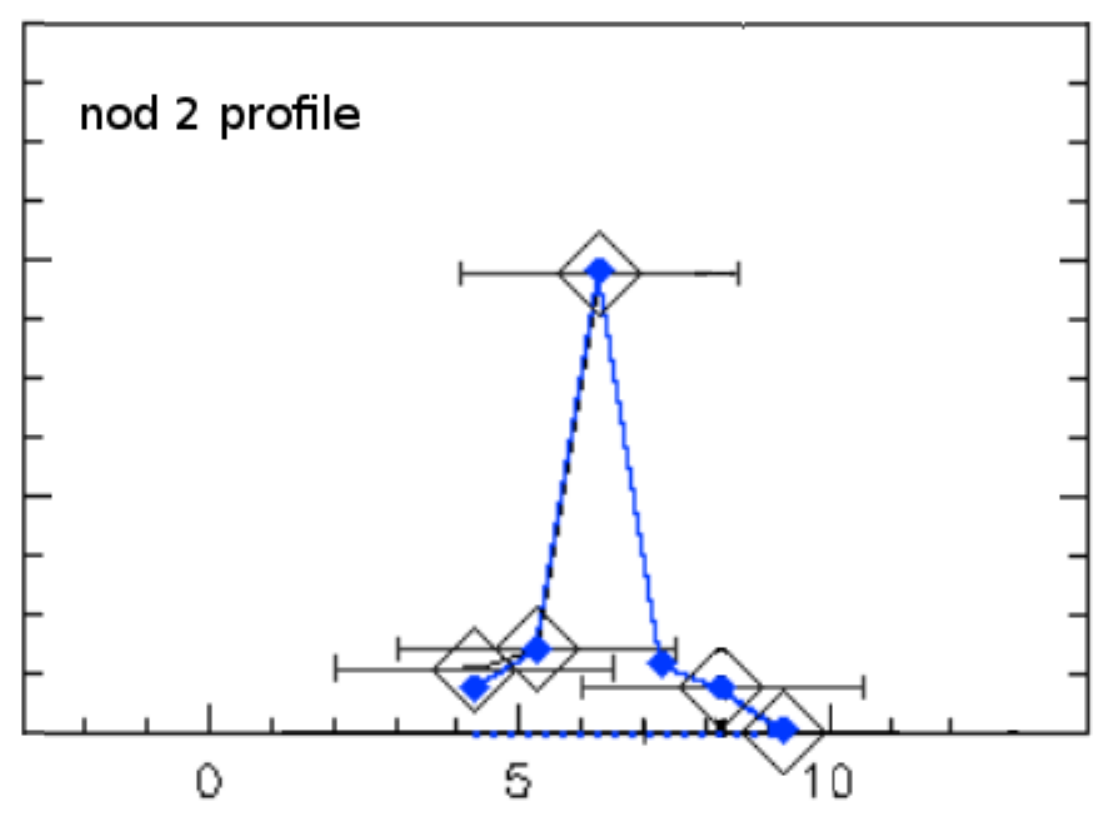}
\includegraphics*[angle=0,width=8cm,height=3.7cm,trim=0 15 0 0]{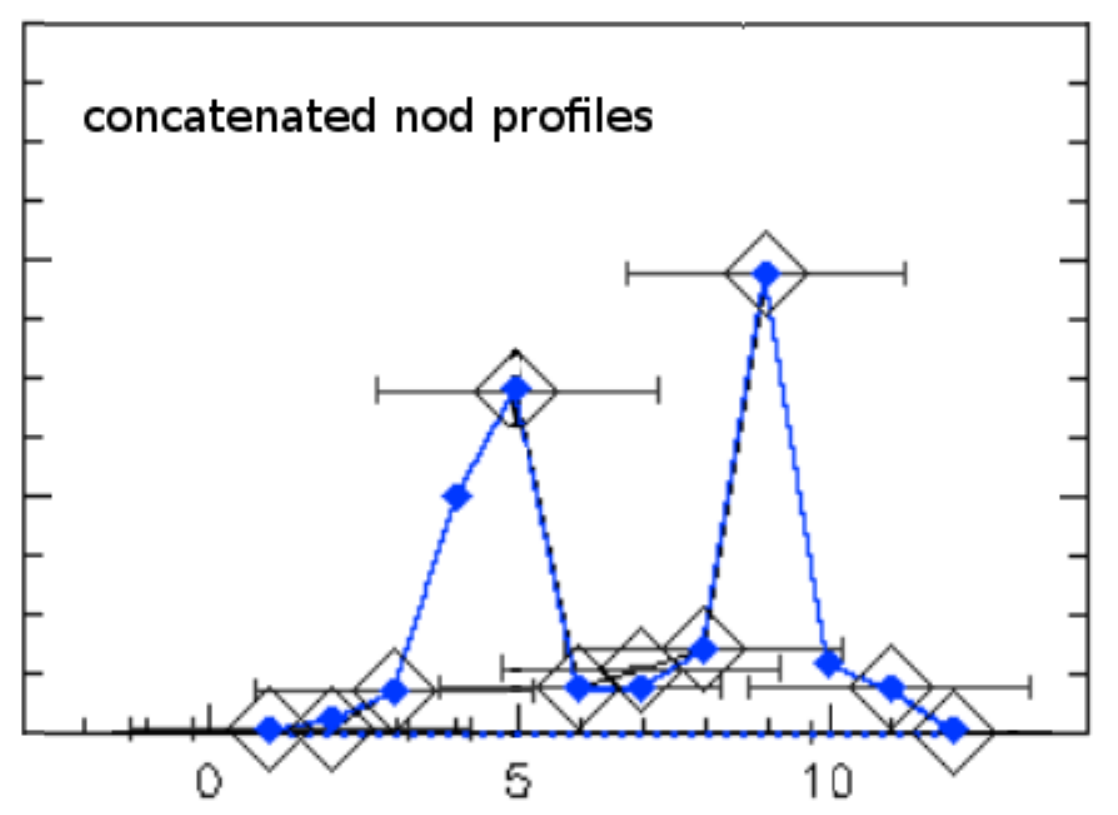}\hskip 1.5em
\includegraphics*[angle=0,width=8cm,height=4.5cm,trim=0 5 0 0]{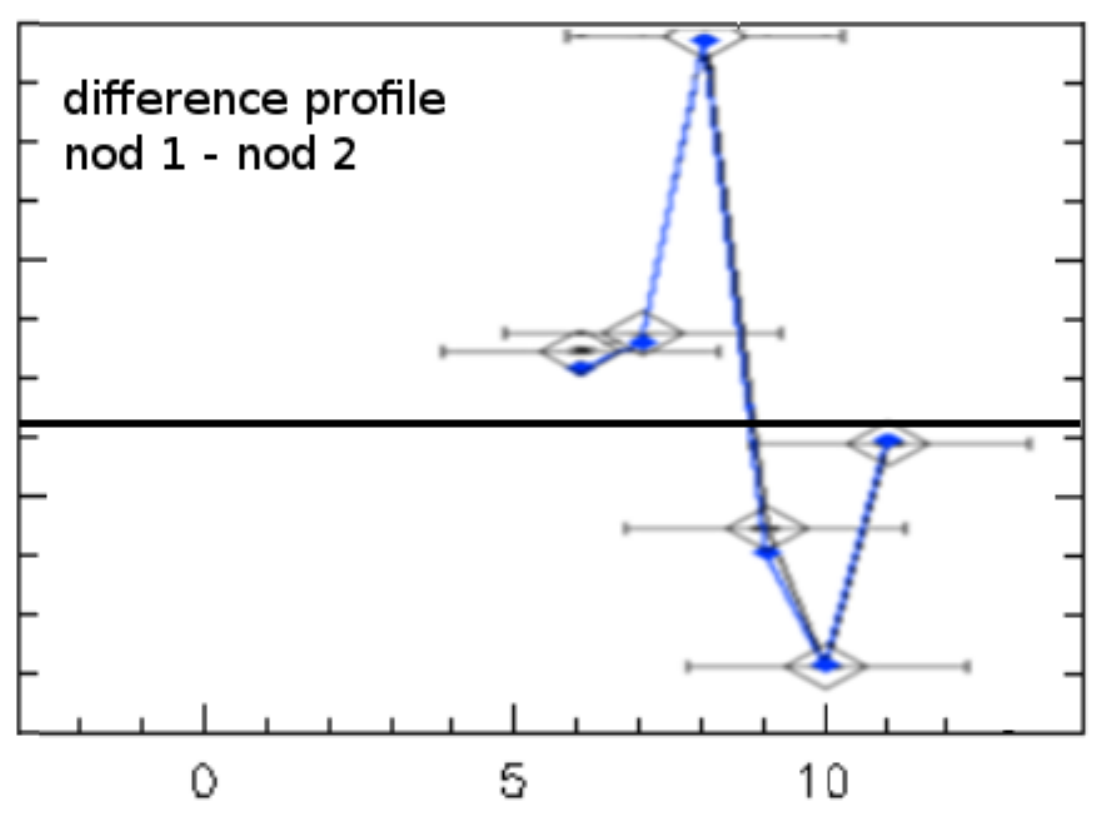}
\caption{Optimal extraction methods for various image combinations. Plots are shown for a given row of a given order (i.e., corresponding to one wavelength element). Data (large diamonds) is plotted with an arbitrary flux density scale as a function of the pixel sequence number along the row. In the nod 1 profile, one data point is missing from the profile, but the remaining points allow for a reliable scaling. In the concatenated nod profiles, the profiles are combined into a single profile and the PSF scaling factor is the same for both nods  (although it appears different because of the different sampling). }
\label{fig:flavors}
\end{figure}

\subsubsection{Differential method}\label{sec:diff}

Another method consists in subtracting the two nod images from each other (Fig.\,\ref{fig:flavors}). This subtraction is a reliable way of correcting low-level rogue pixels whose responsivity remains abnormal over a typical observation timescale (Fig.\,\ref{fig:imdiff3}). Since such rogue pixels are numerous in the detector images, a cleaning algorithm is rendered ineffective and in fact more harmful than useful. 

For the differential optimal extraction, the two nod images are subtracted from each other (uncertainties being propagated). The subtraction produces a differential source profile which is scaled to the model in the same way as the regular method, except that the model is itself a differential super-sampled PSF. The latter is created from two normal super-sampled PSFs shifted in positions and inverted in flux. 

The differential method is adapted for the extraction of unresolved sources, even when the latter are entangled with extended emission. As long as the extended emission is physical, the difference between the two nod images effectively removes the background and produces the spectrum of the unresolved source. 
As explained in Sect.\,\ref{sec:opt_extended}, the differential optimal extraction method is not adapted for partially-extended sources.

\section{Choosing the best extraction method}\label{sec:opt_extended}

The best method for an unresolved source is undeniably the differential method since it removes the background emission if present, and since it removes the low-level rogue pixels. If the source is not a pure point-like source, the differential method fails and underestimates the flux. 

For sources that are almost point-like source, the regular optimal extraction method (i.e., simultaneous extraction of the two nod images) provides a good alternative to the full aperture extraction, with a relatively larger signal-to-noise ratio and with a way of removing the background emission. Taking the extreme example of a source illuminating the aperture uniformly, the difference between the two nod images simply results in a image with pixel values around $0$, with a dispersion corresponding to the S/N of the observation. In such a case, the spectrum will only show noise. In the regular extraction of the same source, the PSF will be scaled to fit the flat profile as best as possible, producing a spectrum with good S/N (though the flux calibration cannot be adequately performed, resulting in a wrong absolute flux and possibly a wrong overall spectral shape).

Both methods of optimal extraction (regular and differential) are not adequate for significantly extended sources because the source spatial profile cannot be modeled with a PSF. In such cases, the full-aperture extraction is the best method. Following the recent improvements of the low-resolution pipeline (see Appendix\,\ref{app:lowres}), the best extraction method is chosen automatically between full aperture for extended sources and optimal extraction for unresolved sources (Sect.\,\ref{sec:opt_extended}). The source extent along the cross-dispersion direction is calculated by comparing the width of the source spatial profile to the FWHM of the PSF. Note that, like for low-resolution observations, we assume that the source is either unresolved or extended for the entire wavelength range. An automatic determination of a wavelength-dependent spatial extent (and the appropriate flux calibration) can only be performed in specific cases \citep{DiazSantos10,DiazSantos11}. The recommended choice of the best extraction method is accompanied in the website with an explanation and with a link to the spatial profile plot.

When the detection level is lower than a certain empirical threshold, the source is assumed to lie at the nod position and optimal extraction is chosen. The reason for this is that if the source is not detected, full aperture extraction will simply add noise while optimal extraction can provide a useful upper limit on the flux, maybe even detecting some spectral feature. For low resolution, the spatial extent determination is relatively more robust because the sources are small compared to the slit length, so the application of a PSF is more reliable. The spatial extent derived from low-resolution observations, when available, is therefore used instead of the one derived from the high-resolution observation to determine the best extraction method. 

\begin{figure}[h!] 
\includegraphics*[angle=0,width=8cm,height=3.5cm,trim=0 1.4cm 0 0]{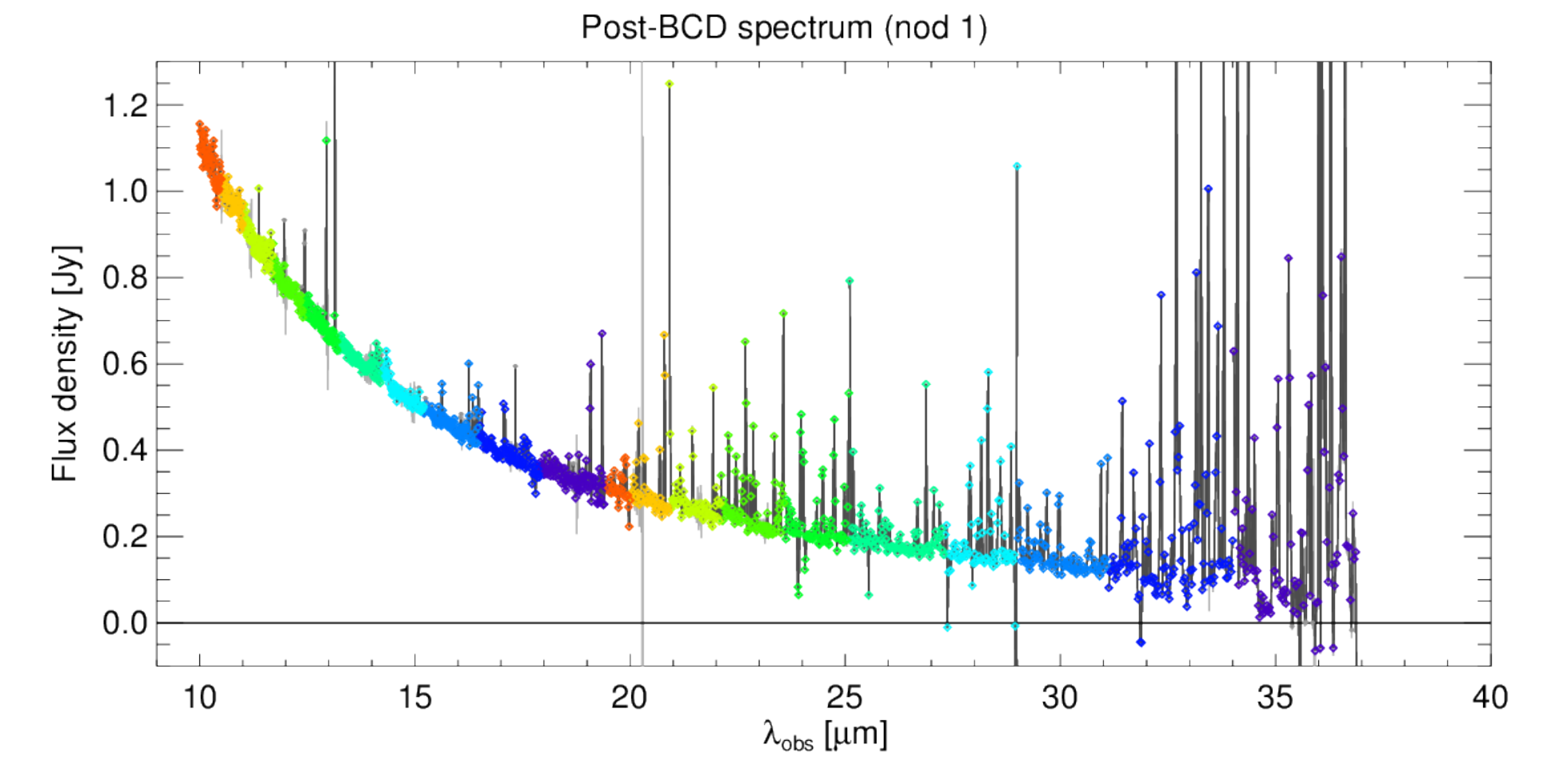}
\includegraphics*[angle=0,width=8cm,height=3.5cm,trim=0 1.4cm 0 0]{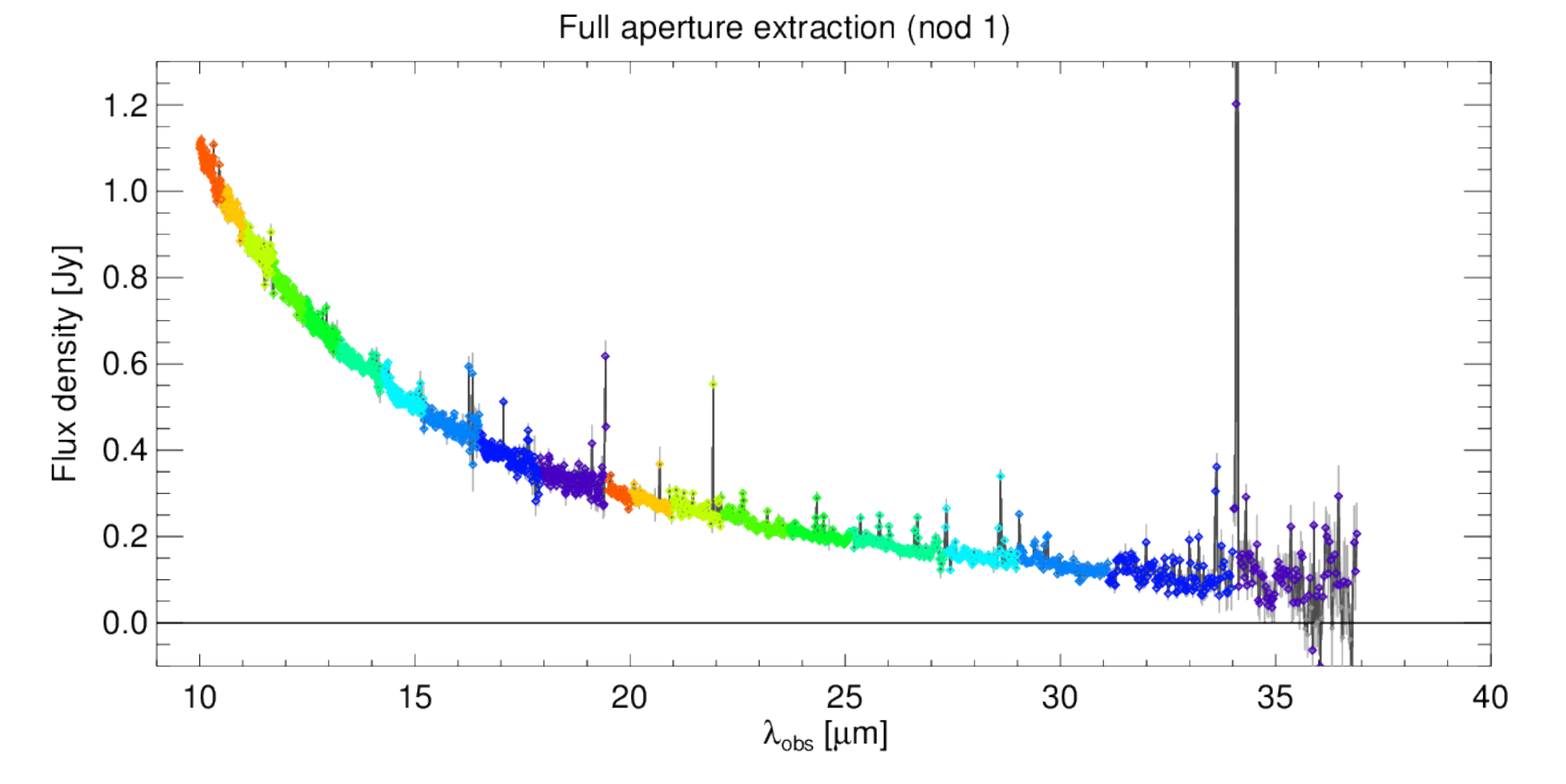}
\includegraphics*[angle=0,width=8cm,height=3.5cm,trim=0 1.4cm 0 0]{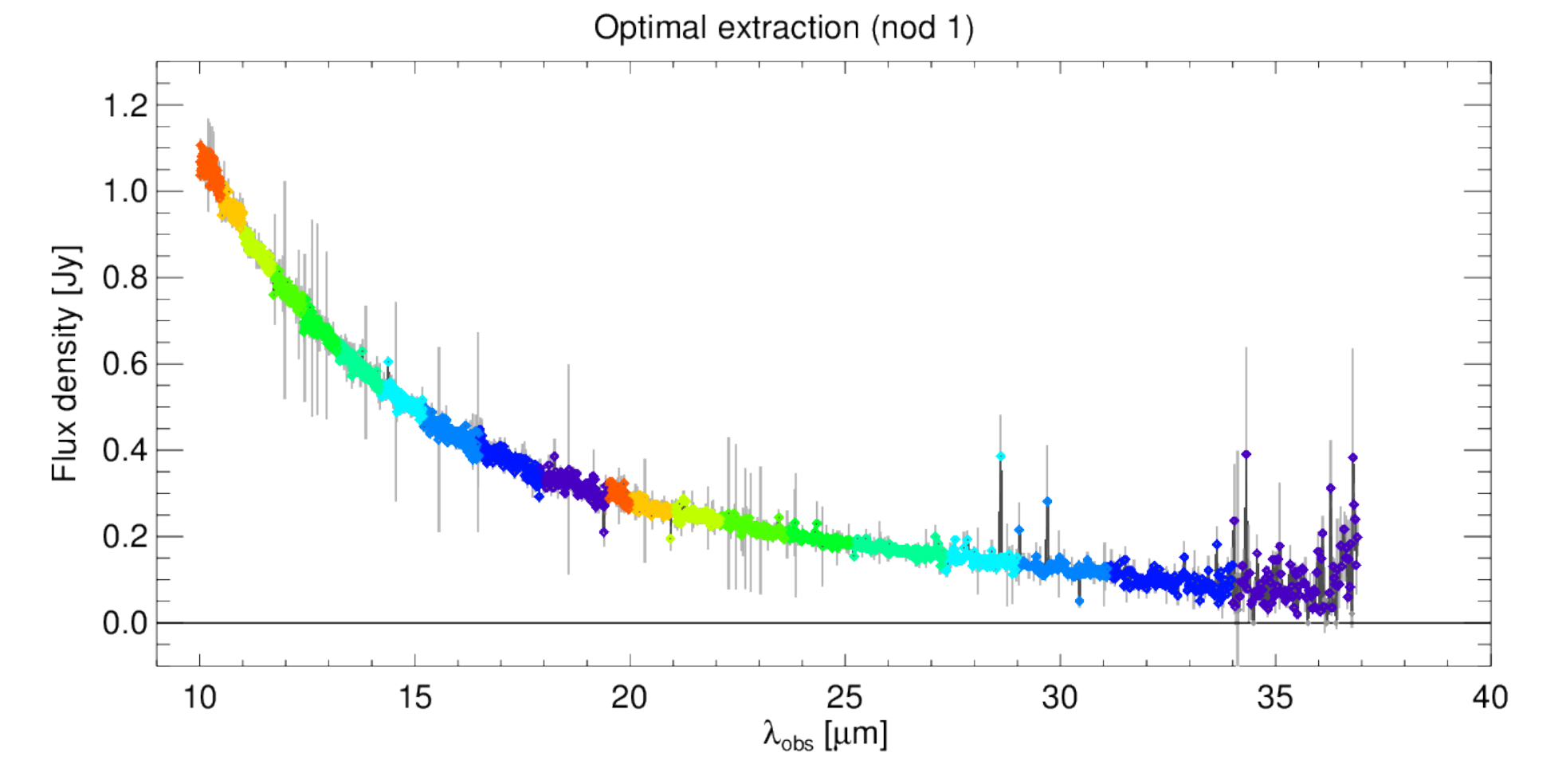}
\includegraphics*[angle=0,width=8cm,height=3.5cm,trim=0 1.4cm 0 0]{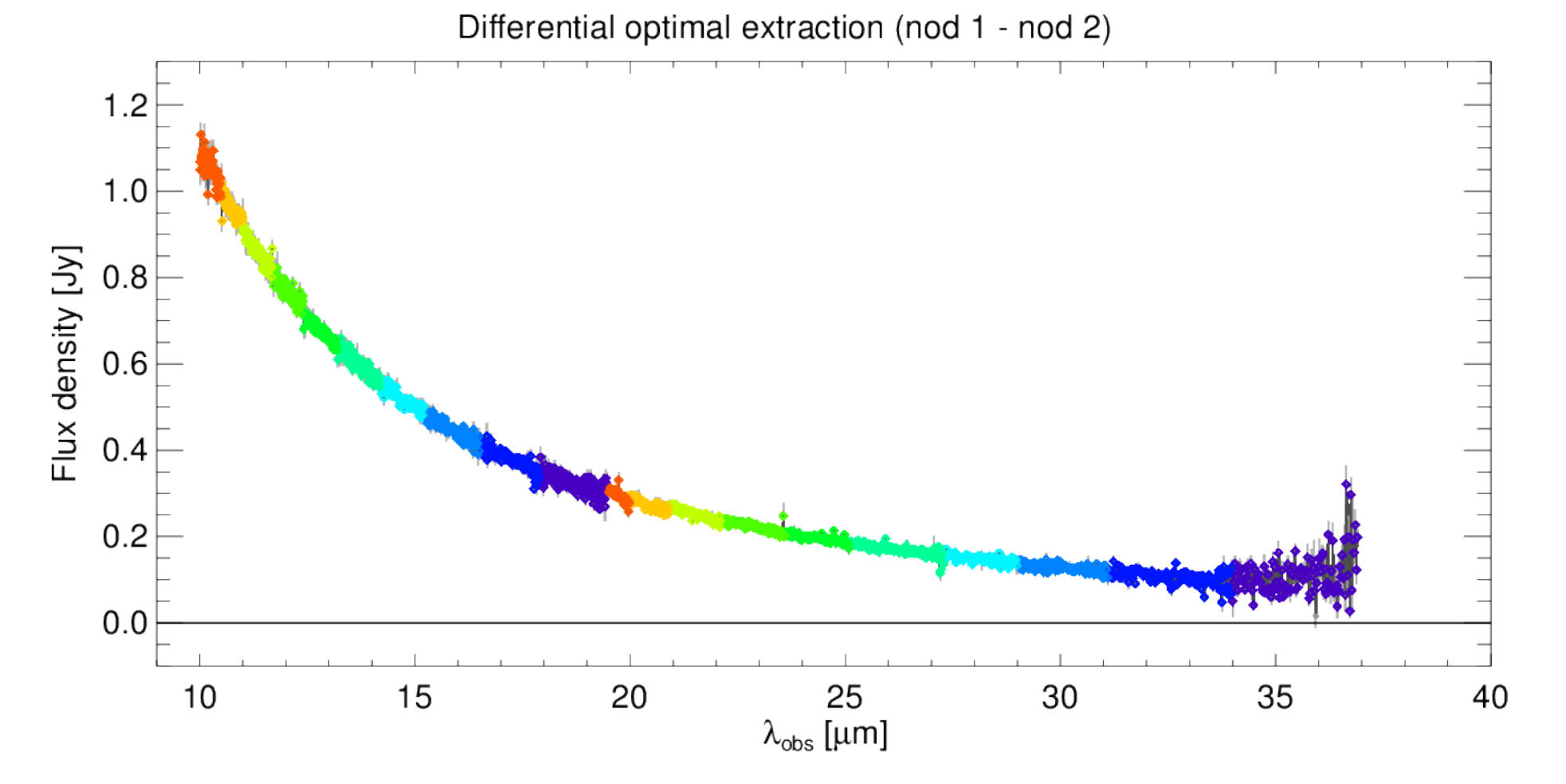}
\includegraphics*[angle=0,width=8cm,height=4.cm,trim=0 0 0 0]{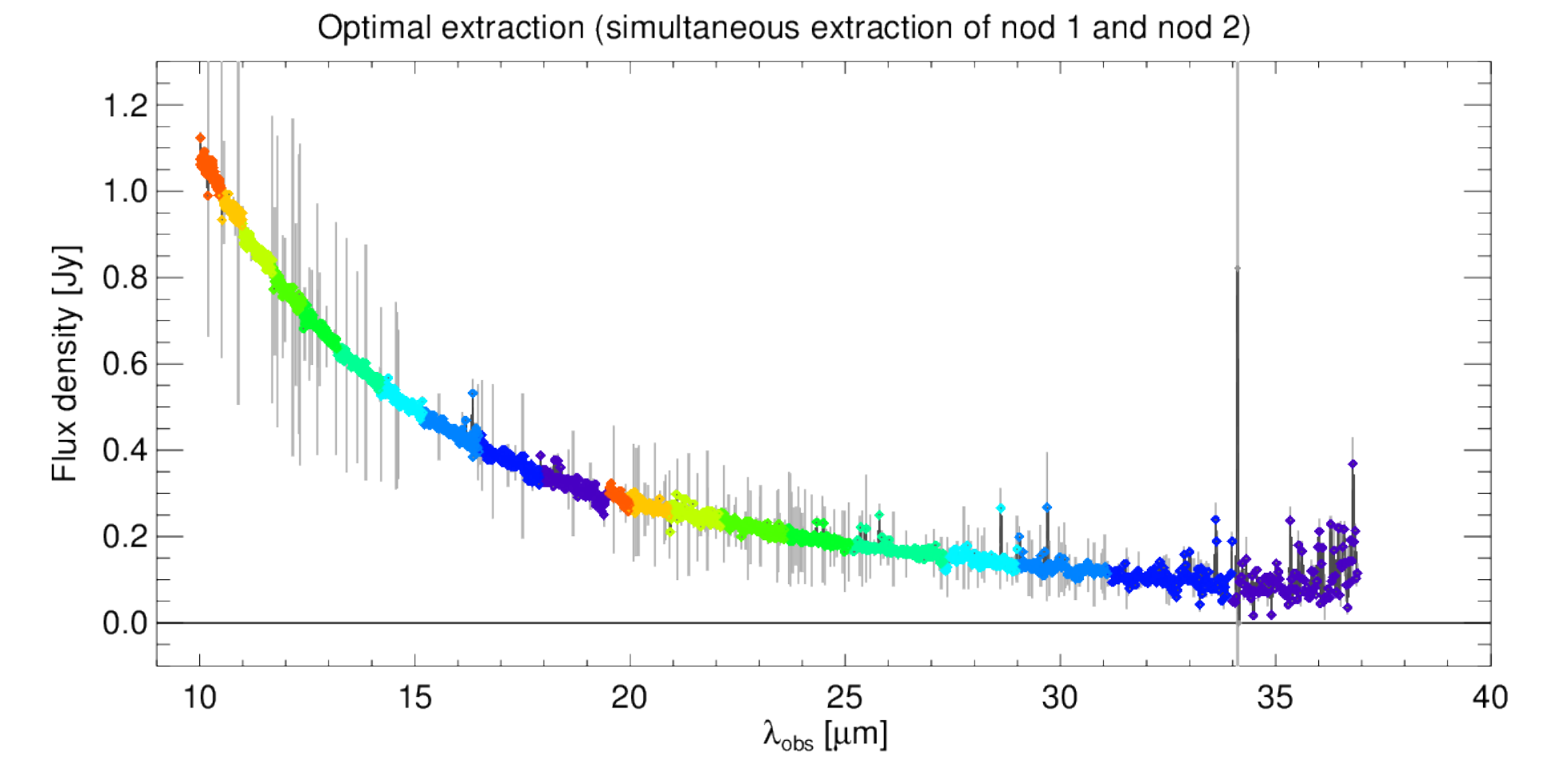}
\caption{SH+LH spectrum of the bright unresolved star $\xi$\,Dra (AORkey 13349632). The best spectrum is provided by the differential optimal extraction. RMS errors are shown in dark gray and systematic errors in light gray. The various colors indicate different spectral orders. The flux density scale is the same for all plots to illustrate the improvement in the RMS noise.
}
\label{fig:spectra1}
\end{figure}

\begin{figure}[h!]
\includegraphics*[angle=0,width=8cm,height=3.5cm,trim=0 1.4cm 0 0]{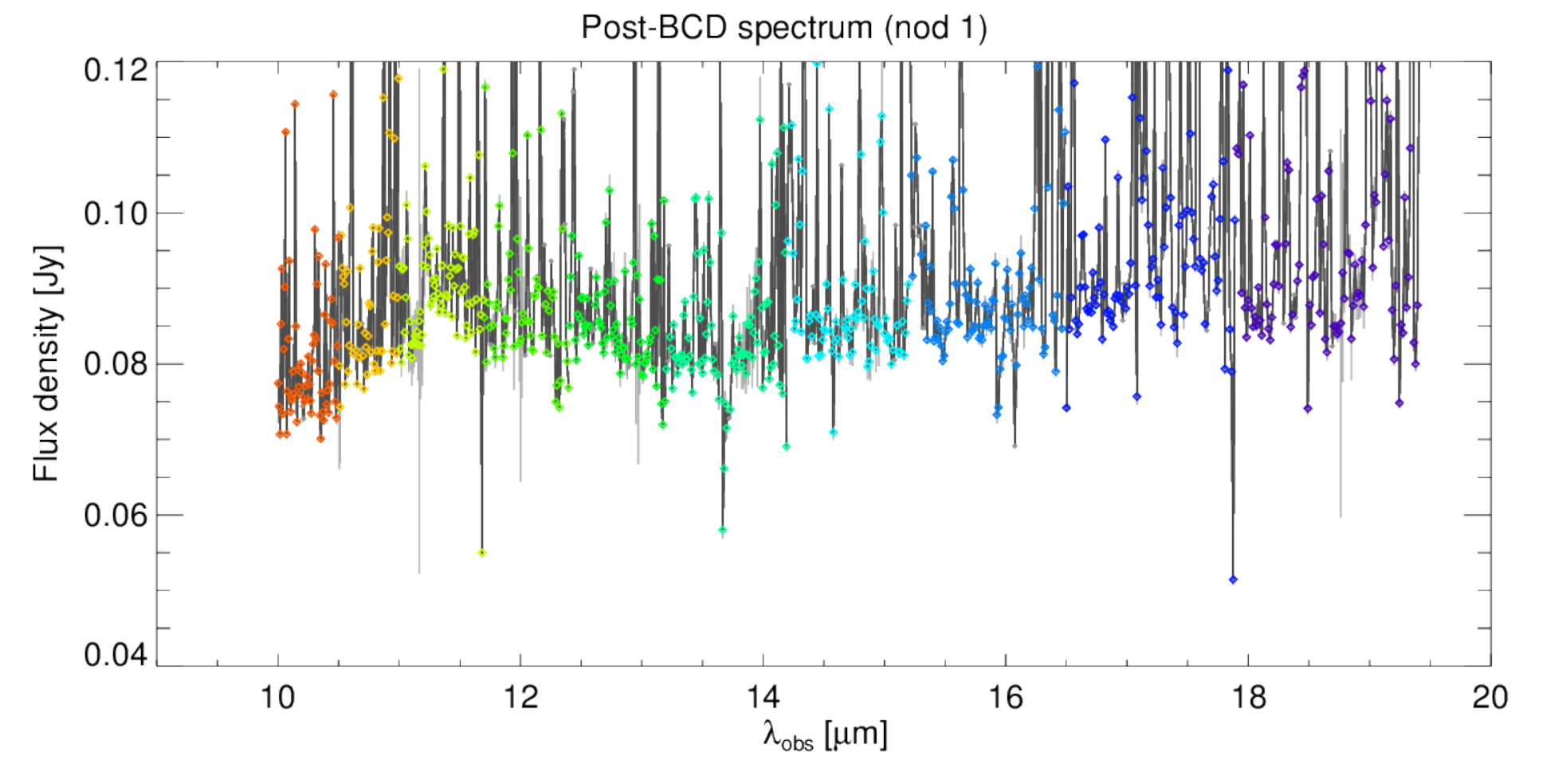}
\includegraphics*[angle=0,width=8cm,height=3.5cm,trim=0 1.4cm 0 0]{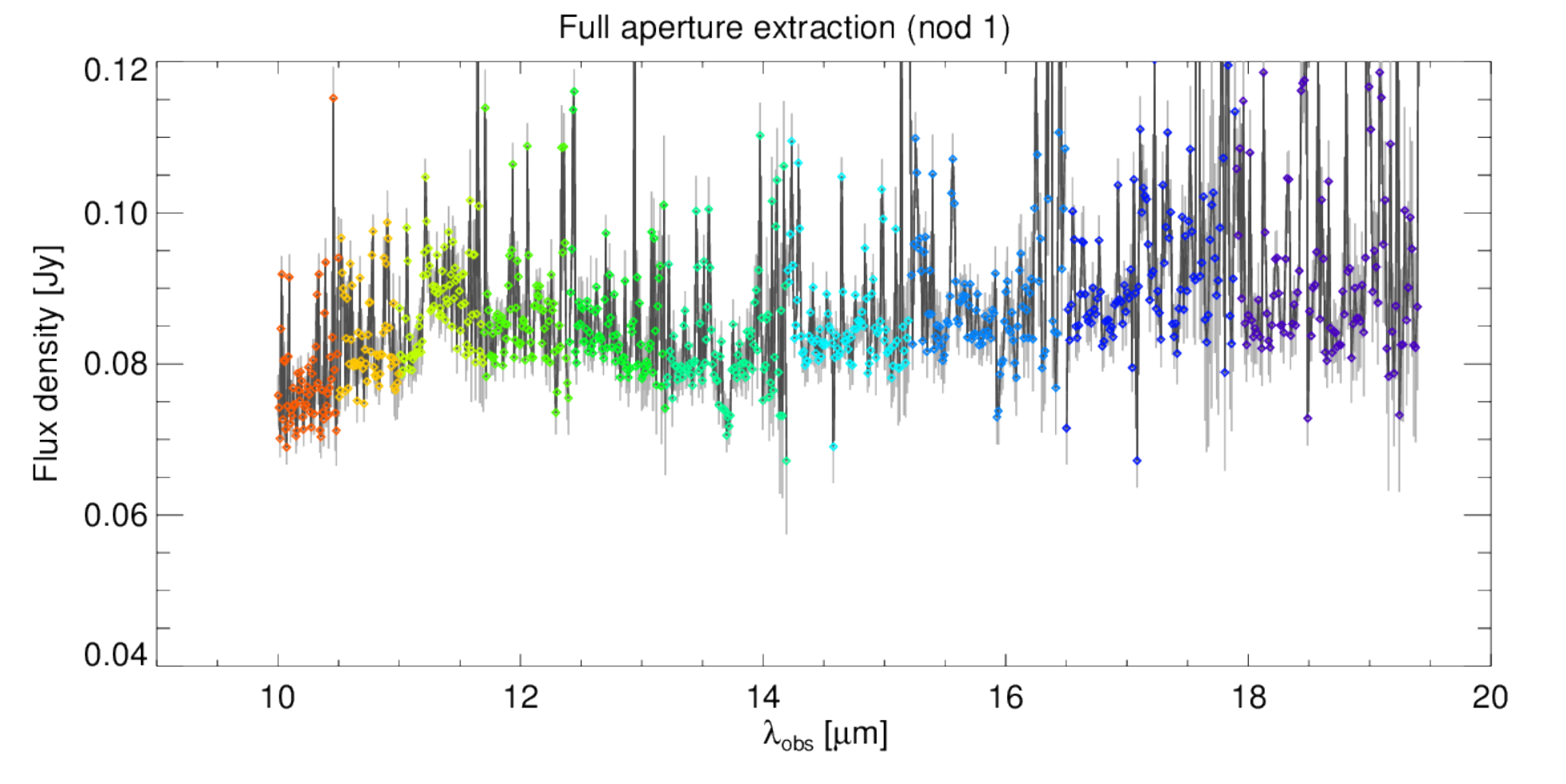}
\includegraphics*[angle=0,width=8cm,height=3.5cm,trim=0 1.4cm 0 0]{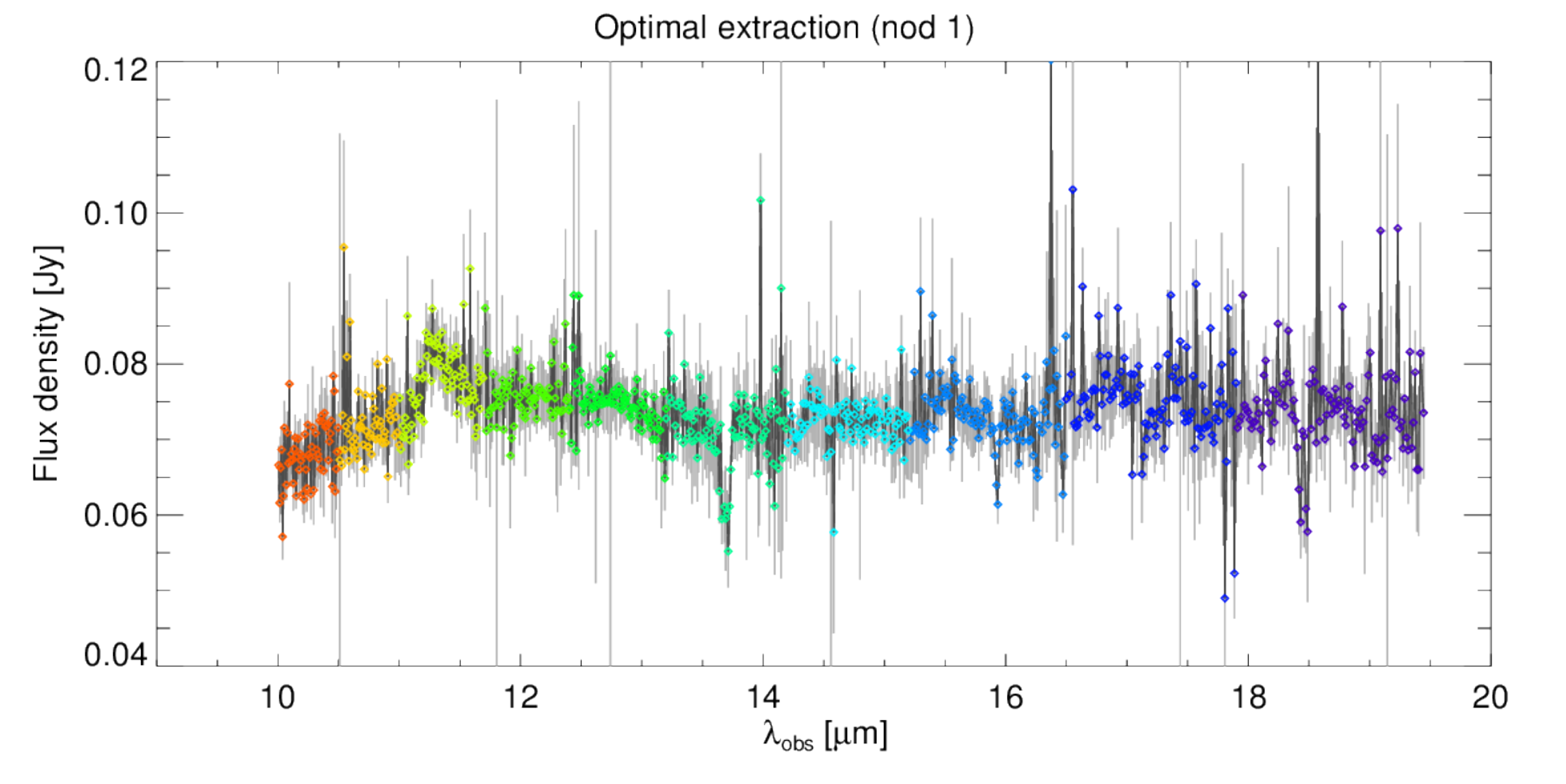}
\includegraphics*[angle=0,width=8cm,height=3.5cm,trim=0 1.4cm 0 0]{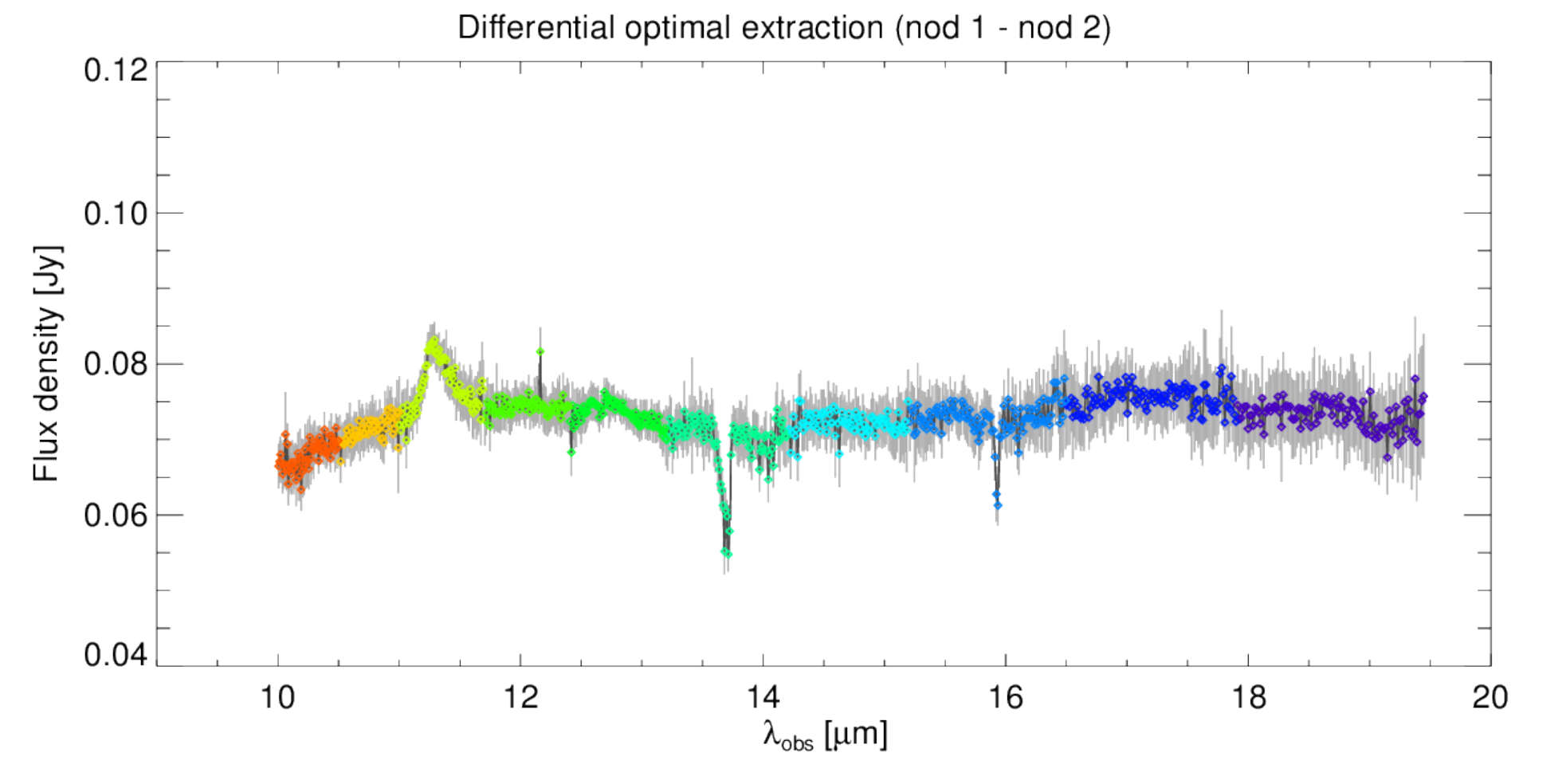}
\includegraphics*[angle=0,width=8cm,height=4.cm,trim=0 0 0 0]{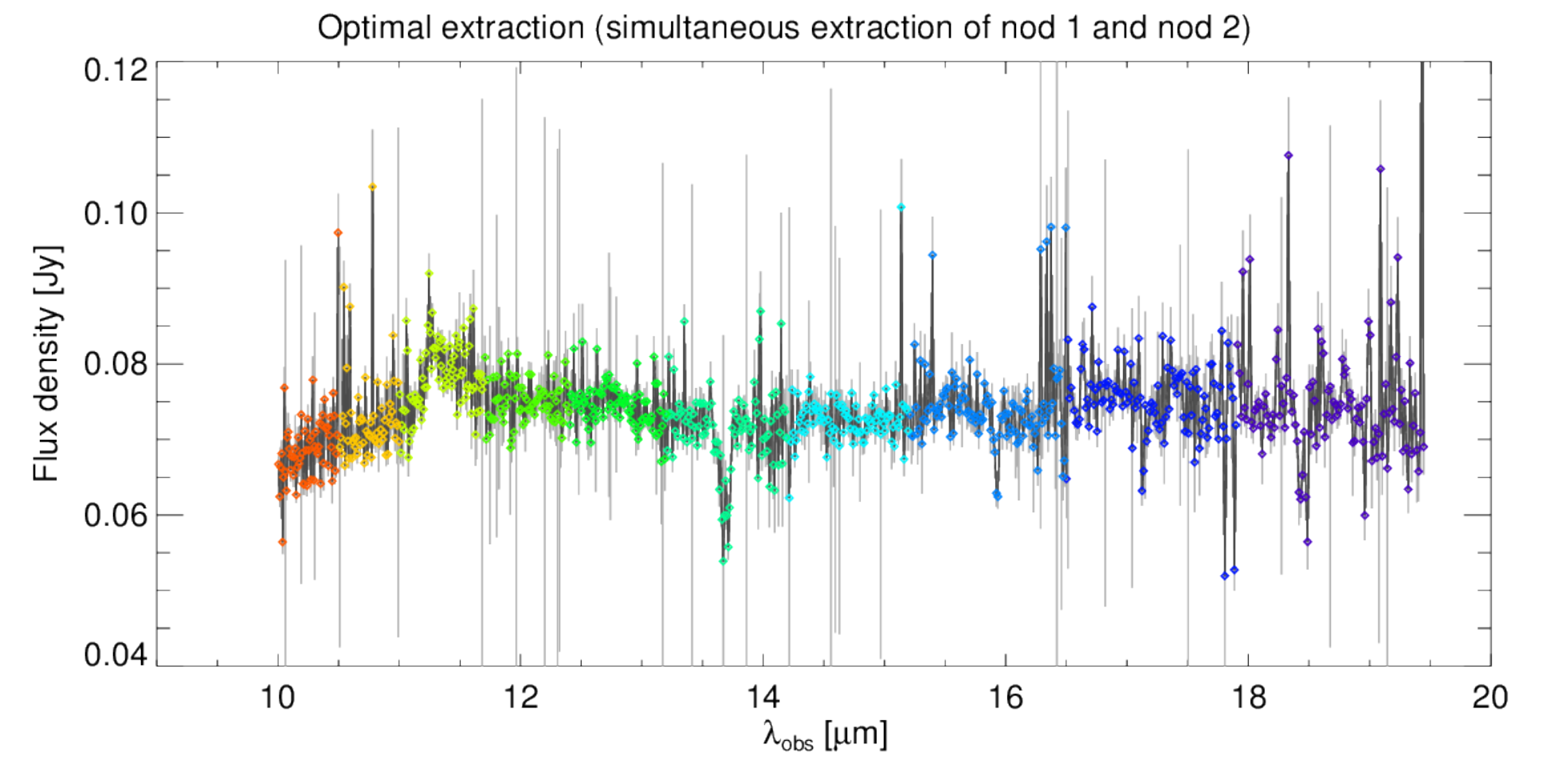}
\caption{SH spectrum of the relatively faint post-AGB star MSX SMC029 (AORkey 25646848; \citealt{Kraemer06}). See Fig.\,\ref{fig:spectra1} for the plot description. The emission feature at $11.3$\mic\ and absorption features at $\approx13.5$\mic\ and $\approx16$\mic\ are real and best seen in the optimal extraction versions. 
}
\label{fig:spectra2}
\end{figure}

\begin{figure}[h!]
\includegraphics*[angle=0,width=8cm,height=3.5cm,trim=0 1.4cm 0 0]{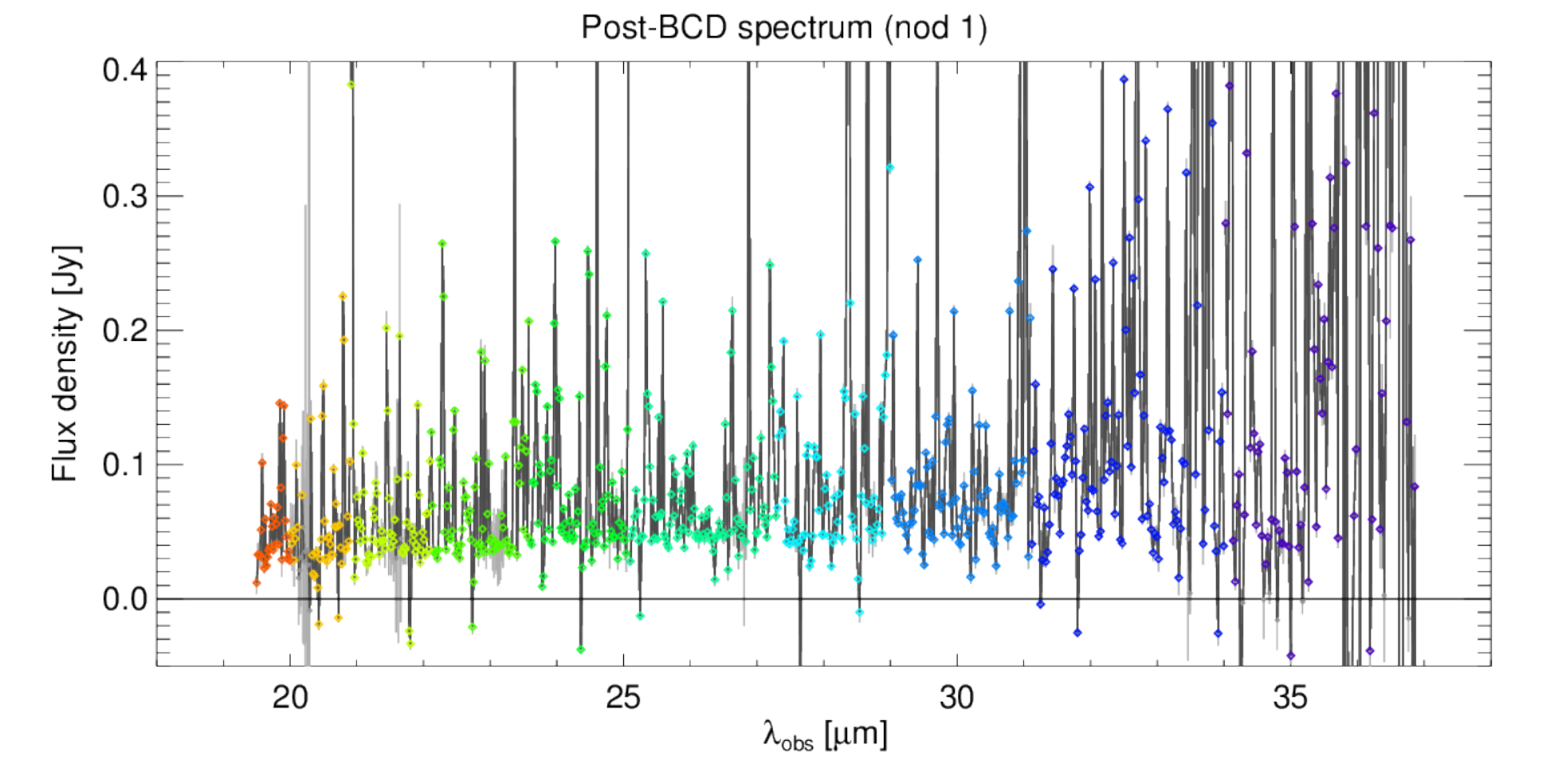}
\includegraphics*[angle=0,width=8cm,height=3.5cm,trim=0 1.4cm 0 00]{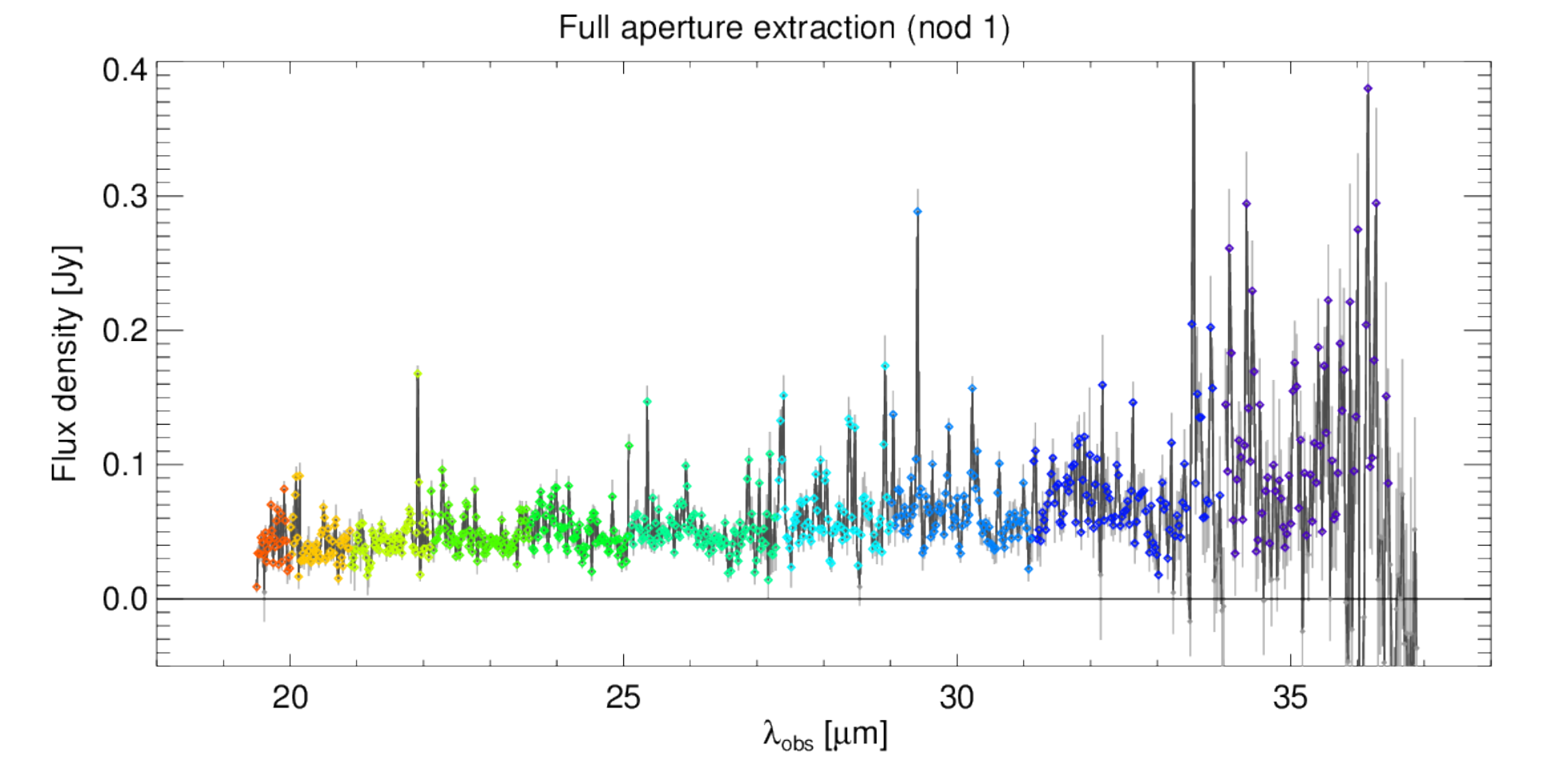}
\includegraphics*[angle=0,width=8cm,height=3.5cm,trim=0 1.4cm 0 0]{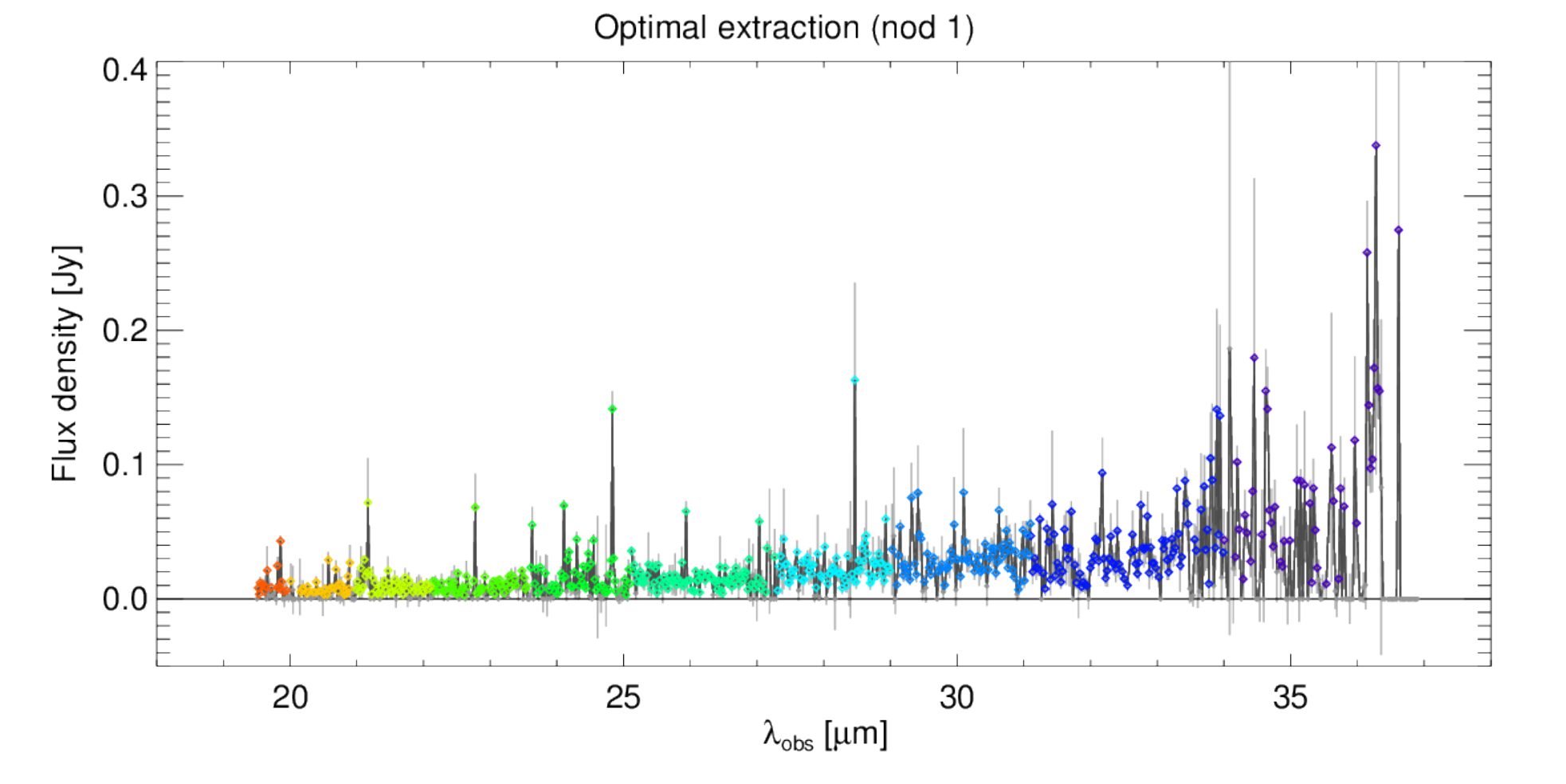}
\includegraphics*[angle=0,width=8cm,height=3.5cm,trim=0 1.4cm 0 0]{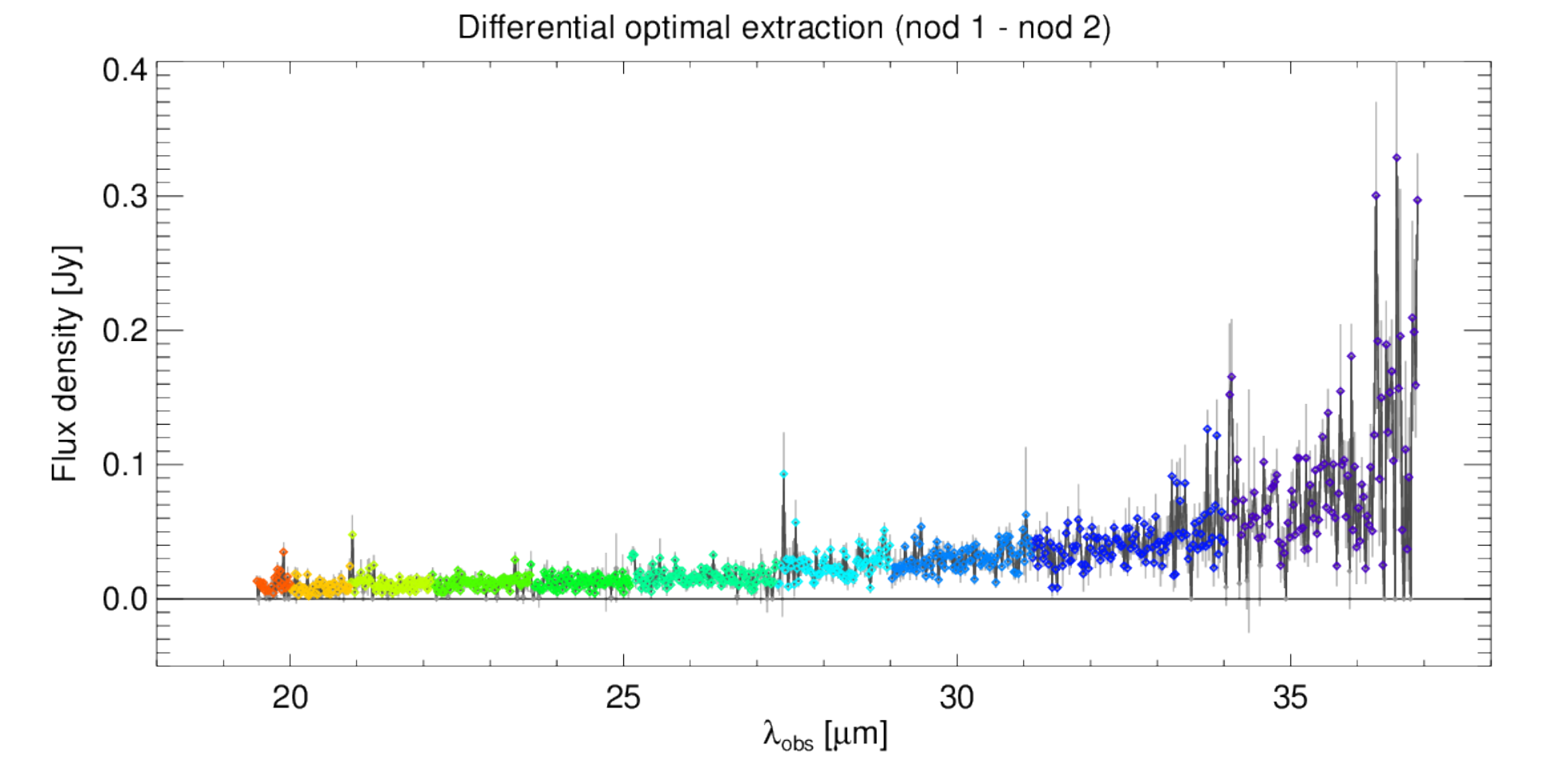}
\includegraphics*[angle=0,width=8cm,height=4.cm,trim=0 0 0 0]{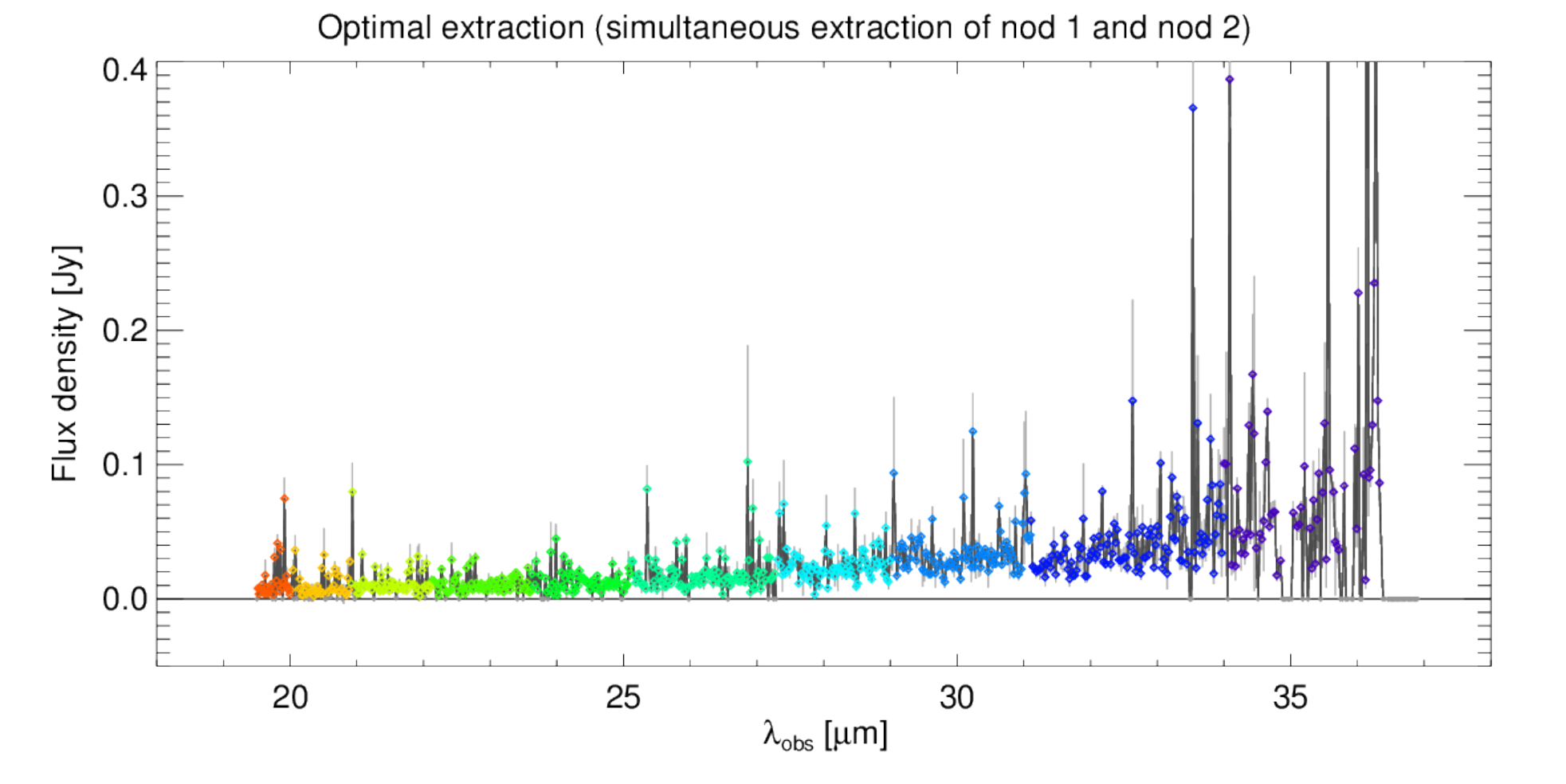}
\caption{LH spectrum of the faint unresolved galaxy 2MASX J10394598+6531034 (AORkey 33357568). See Fig.\,\ref{fig:spectra1} for the plot description. Note that the source has no detectable emission lines. The flux scale is the same for all plots, illustrating how optimal extraction removes the background emission that is included in full aperture extractions.  }
\label{fig:spectra3}
\end{figure}

Examples of spectra extracted with full aperture and optimal extraction are shown in Figs.\,\ref{fig:spectra1}, \,\ref{fig:spectra2}, and  \,\ref{fig:spectra3}, corresponding to sources of various brightness. As can be seen from these plots, the differential optimal extraction generally provides the cleanest spectrum and best overall S/N for an unresolved source because low level rogue pixels are removed, but this extraction yields incorrect flux densities if there is any extension of the source. The regular optimal extraction (simultaneous extraction of the two nod images) also provides a significant improvement in S/N over full-aperture extraction and is valid for marginally extended sources. For extended sources, the CASSIS full-aperture extraction provides a significant improvement over the post-BCD spectrum.  It should be emphasized that optimal extraction removes the extended background and therefore yields a spectrum with overall flux values smaller than the full aperture extraction. 

The choice of ``best'' extraction is dependent, therefore, on real source extent beyond the spatial profile of the PSF. Because the measure of extent is uncertain, the current version of CASSIS provides simultaneous nod extraction as the default optimal extraction for comparison to the full-aperture extraction. However, all choices are given in CASSIS so that the user can compare and select a final spectrum based on the best estimate of source extent.

\section{Post-processing of the spectra}\label{sec:specsteps}

\subsection{Fringes removal}\label{sec:fringes}

Fringes originate between plane-parallel surfaces in the light path of the instrument. The surfaces act as Fabry-P\'erot etalons, each of which can add unique fringe components to the source signal. While the LL1 fringes are believed to be the result of a filter delamination discovered prior to launch, the SH and LH fringes originate from the detector substrates and probably also a filter in the IRS \citep{Lahuis07PhD,Lahuis03}. 
Fringes are removed for any extraction type (optimal and full aperture) with the \texttt{IRSFRINGE} tool\footnote{\url{http://irsa.ipac.caltech.edu/data/SPITZER/docs/dataanalysistools/tools/irsfringe/}}. The default SH and LH settings in \texttt{IRSFRINGE} are chosen to identify fringes. Fringes are looked for in each order individually. 

The flat-fielded (BCD) images already include a fringe correction, although residual fringes may remain. Such a correction relies on the assumption that the fringe phase, and to a lesser extent the amplitude, does not vary greatly between different observations. We decided to remove the fringes in the unflat-fielded images and correct for the flat-field within the flux calibration step (Sect\,\ref{sec:calib}). 

\subsection{Spectral order overlaps}

The spectra from consecutive spectral orders in high-resolution observations sometimes overlap significantly. In the current version of the atlas, we do not attempt to combine the spectra in overlap regions but simply choose the order cutoffs to create a continuous spectrum with no overlaps. We have investigated a large number of sources to determine empirically the best cutoffs for each spectral order. The current values ensure that the order providing the best S/N is chosen for any particular wavelength range. Untrimmed spectra are available by request.

\subsection{Flux calibration}\label{sec:calib}

The flux calibration for optimal extraction and full aperture extraction was performed using a relative spectral response function (RSRF) calculated from $76$ observations of $\xi$\,Dra (including $20$ before campaign $25$, .i.e., before the LH detector bias voltage was set to its final value). The theoretical templates and calibration method is tied to the low-resolution spectral calibration by \cite{Sloan14} since $\xi$\,Dra was observed in both low- and high-resolution. 

Separate RSRFs were created for optimal extraction (regular and differential methods) and full aperture extraction at both nod positions. This means that each extraction method has been empirically calibrated using a point source, so flux calibration is precise only for point sources. The calibration uncertainty is a systematic error that depends on wavelength.

\section{Dedicated background offset observations}\label{sec:bgsub}

Subtracting an offset background image can be a useful way to cancel out pixels with responsivity variations that cannot be calibrated reliably (low-level rogue pixels). Since the IRS high resolution modules were designed primarily to study emission lines, correctly subtracting the underlying continuum was not initially considered sufficiently important to double recommended observing times by taking a separate background spectrum.  Later in the \textit{Spitzer} mission, rogue pixels developed from cosmic ray damage by unexpectedly strong solar flares.  Only after that time were observers advised to include offset pointings for high resolution observations.  As a result, a significant fraction of observations do not have specific offsets (or in some cases the offset images were not observed with the same exposure time as the science images). 

For these reasons and also because the identification of offset backgrounds \textit{a posteriori} is not straightforward, as there was no standard way of designing offset observations, we have chosen not to utilize any available background observations in the present version of the pipeline. For unresolved sources embedded in large-scale emission, both optimal extraction methods (Sect.\,\ref{sec:optext_flavors}) effectively disentangle both components. For extended sources, full-aperture extraction is the best method, and no background can be removed except a dedicated offset observation. As explained in Sect\,\ref{sec:fullap}, if the observation ID of the dedicated background observation is known, users can download the spectrum of the dedicated background observation and subtract it from the science source spectrum. The use of dedicated offset backgrounds for removal at the image level will be investigated in the future for CASSIS.

\section{Summary}

We present the high-resolution spectral pipeline for the \textit{Spitzer}/IRS instrument. The corresponding atlas is available online at \url{http://cassis.sirtf.com}, complementing the existing atlas for low-resolution data presented in L11.

High-resolution modules on the IRS are particularly plagued by cosmic ray hits and a particular attention was given to the exposure combination and image cleaning to remove as many bad pixels as possible. The pipeline produces a full-aperture extraction for extended sources. Unresolved sources are extracted with an optimal extraction using a super-sampled PSF, the latter created for the first time for the IRS high-resolution modules. Two optimal extraction methods are considered, (1) a method extracting the two nod images simultaneously and removing large-scale emission that may be instrumental or physical, (2) a method extracting the difference of the nod images, allowing a complete removal of any large-scale emission and of some instrumental artifacts.

\begin{acknowledgements}
We are grateful to F.\ Lahuis for fruitful discussions on the high-resolution optimal extraction techniques. 
We wish to thank again the people who contributed to the data reduction efforts over the IRS mission. Former ISC members are especially acknowledged (in particular D.\,Devost, D.\,Levitan, D.\,Whelan, K.\,Uchida, J.D.\,Smith, E.\,Furlan, M.\,Devost, Y.\,Wu, L.\,Hao, B.\,Brandl, S.J.U.\ Higdon, P.\ Hall) for their work on the SMART software and for the development of reduction techniques. Moreover, our colleagues in Rochester (M.\,McCLure, C.\,Tayrien, I.\,Remming, D.\,Watson, and W.\,Forrest) played an important role in bringing additional and essential improvements to the data reduction used in CASSIS. 
This research has made use of the NASA/IPAC Extragalactic Database (NED) which is operated by the Jet Propulsion Laboratory, California Institute of Technology, under contract with the National Aeronautics and Space Administration. 
This research has made use of the SIMBAD database, operated at CDS, Strasbourg, France. 
This research was conducted with support from the NASA Astrophysics and Data Analysis Program (Grant NNX13AE66G)
\end{acknowledgements}

\appendix
\appendixpage

\section{Low-resolution pipeline updates}\label{app:lowres}

Since publication of L11, several updates have been made to the CASSIS low-resolution pipeline. The current version is labelled as ``LR7''. The spectra now available through the CASSIS website (\url{http://cassis.sirtf.com}) include the following improvements:
\begin{itemize}
\item The best extraction method is chosen automatically between optimal extraction and tapered column extraction (integrating the flux within a spatial window whose width scales with wavelength), based on the source spatial extent. Optimal extraction is best used for unresolved sources, while tapered column extraction is adapted for partially-extended sources. The default spectrum shown on the main result page and the default products reflect the automatic choice between the extraction methods. The alternative method can still be accessed through the options.
\item Background subtraction for low-resolution observations was performed either by removing the detector image(s) corresponding to the other spectral order ``by-order'') or to the other nod (``by-nod''). The presence of a contaminating source in the nominal image and in the background image(s) is critical to constrain what background subtraction method is eventually used (between by-order, by-nod, or no subtraction at all). The parameters were adjusted so that a contamination is identified as such only when it affects significantly the source spectrum. 
\item Tapered column extractions in v4 were presented with the best background subtraction based on diagnostics drawn from the optimal extraction algorithm (accounting for the presence of contaminating sources). If the source is too extended, however, it becomes impossible to disentangle the ``positive'' and ``negative'' peaks in the differential profile and the by-nod background subtraction is not reliable. For tapered column extractions of partially-extended sources, the subtraction by-order is now preferred, unless there is a contaminating source in the other order background, in which case no by-nod or by-order subtraction can be performed, resulting in what is referred to as ``in situ'' local background removal, which removes only the baseline to the spatial profile as opposed to removing a 2D background image.
\item The by-order (and to a lesser extent by-nod) subtraction sometimes resulted in a significant residual of the extended background emission. This mostly affects very faint sources (typically $\lesssim1$\,mJy) for which the source flux is much smaller than the \textit{difference} of the background emission between two order (or two nods). The residual emission is now removed prior to extraction of the source profile. 
\item The latest and final version of the BCD calibration is used (S18.18.0).
\item Various improvements were made for the website, with in particular the ability to overlay the slits on archival images. 
\end{itemize}

\bibliography{/home/vleboute/Workplace/Cenva/TexStyle/mybib}

\end{document}